\documentclass[12pt]{iopart}

\usepackage{cite}
\usepackage{iopams}
\usepackage{graphicx}
\usepackage{epstopdf}
\usepackage[ruled]{algorithm2e}
\usepackage{url}
\usepackage{color}
\def\n{{\mathbf n}}
\def\f{\frac}
\def\q{\quad}
\def\na{\nabla}
\def\Om{\Omega}
\def\mE{{\mathcal E}}
\def\p{\partial}
\def\R{{\mathbb R}}

\begin{document}

\title[Monotonicity-Based EIT for Lung Imaging]{Monotonicity-based Electrical Impedance Tomography for Lung Imaging}

\author{Liangdong Zhou$^{1}$,Bastian Harrach$^{2}$, Jin Keun Seo$^{3}$}

\address{$^{1}$ Department of Radiology, Weill Cornell Medical College, New York, NY, USA}
\address{$^{2}$ Institute for Mathematics, Goethe-University Frankfurt, Frankfurt am Main, Germany}
\address{$^{3}$ Department of Computational Science and Engineering, Yonsei University, Seoul, South Korea}
\ead{harrach@math.uni-frankfurt.de}
\vspace{10pt}
\begin{indented}
\item[]January 2018
\end{indented}

\begin{abstract}
This paper presents a monotonicity-based spatiotemporal conductivity imaging method for continuous regional lung monitoring using  electrical impedance tomography (EIT). The EIT data (i.e., the boundary current-voltage data) can be decomposed into pulmonary, cardiac and other parts using their different periodic natures. The time-differential current-voltage operator corresponding to the lung ventilation can be viewed as either semi-positive or semi-negative definite owing to monotonic conductivity changes within the lung regions. We used these monotonicity constraints to improve the quality of lung EIT imaging.  We tested the proposed methods in  numerical simulations, phantom experiments and human experiments.
\end{abstract}

%
\vspace{2pc}
\noindent{\it Keywords}: Electrical impedance tomography, continuous lung monitoring, monotonicity, inverse problem, monotonicity-based regularization
%
%
%
%

\section{Introduction}
Electrical impedance tomography (EIT) has received much recent attention owing to its unique ability to allow  long-term, continuous  monitoring of lung ventilation at the bedside \cite{Adler1996Eng,Adler2009}, which is not possible using other medical imaging techniques such as computerized tomography (CT), magnetic resonance imaging (MRI), ultrasound (US), or single photon emission computed tomography (SPECT). The increasing sophistication of monitoring  intensive-care patients might incorporate the use of  EIT system to guide strategies for  protective  lung  ventilation via the close monitoring of  patient's lung \cite{Meier2008,Putensen2007,Kunst2000,Wolf2013,Bikker2009,Costa2008,Costa2009}. Although EIT cannot compete with CT, MRI or ultrasound in terms of spatial resolution or accuracy \cite{Holder2005book}, its ability to provide long-term, continuous monitoring and portability make it clinically useful.

This paper focuses mainly on the lung EIT. In EIT, multiple surface electrodes are attached to an imaging object to inject currents and measure boundary voltages. At  frequencies below about 250 kHz, the potential induced by the injection current is dictated approximately by the elliptic partial differential equation $\na\cdot(\sigma\na u)=0$ inside the object where $\sigma$ is the time-varying conductivity distribution associated with lung ventilation \cite{Holder2005book,Seo2012book}. The measured  EIT data can be viewed as a boundary current-voltage map from the injection current to the resulting boundary voltage, which is determined mainly by the effective conductivity distribution,  the configuration of the surface electrodes, and the geometry of the imaging object. Lung EIT aims to provide  dynamic images of the time-differential conductivity distribution from the time-differential  current-voltage map while minimizing the forward modeling errors due to such as electrode position, boundary geometry, and uncertainty of the reference conductivity distribution.

Robust EIT reconstruction algorithms  have been sought since the invention of the first EIT devices by Barber and Brown in the early 1980s \cite{Barber1984a,Barber1984b,Adler1996Imag,Oh2008}. However, EIT  is not yet suitable for routine clinical use because of its poor sensitivity \cite{Teschner}. Robust reconstruction may rely on incorporating some strong prior information into the algorithm via regularization. Unfortunately, it seems inadequate to achieve robust time difference EIT reconstructions only with regularization, such as Tikhonov regularization \cite{Seo2012book,Vauhkonen1998} and total variation regularization \cite{Borsic2010}, in spatial domain. In other words, temporal regularization should be used with the prior information in time domain, especially for time difference EIT.

Taking into account of  the fundamental limitation of EIT, some strong prior information about the individual's pulmonary function needs to be incorporated into the analysis along with  the measured data. We know that the time-varying patterns in lung EIT data depend mainly on the pulmonary and cardiac cycles and diaphragm motion.  Given that  the pulmonary and cardiac cycles have different frequencies, we can extract patterns corresponding to the ventilation  from the measured EIT data by eliminating the signals associated with cardiac motion and other processes \cite{Leathard1994,Grant2011}.

Using the extracted ventilation-induced signals, we apply  the characteristic of monotonicity to the regional lung imaging. This is based on the assumption that changes of conductivity distribution induced by lung ventilation are either monotonically non-increasing or monotonically non-decreasing  at any fixed time. This assumption can be understood in terms of  the effective conductivity, which  means the overall conductivity changes induced by ventilation can be regarded as monotonically non-increasing  or monotonically non-decreasing. An increase of conductivity decreases  the voltage measurements in terms of matrix definiteness. In other words, if the conductivity  increases between two time frames, the measured difference data matrix should be negative semi-definite. Conversely,
a conductivity decrease would increase the voltage measurements in terms of matrix definiteness. We observe that
the time-derivative of the current-voltage data associated with ventilation is a non-negative
operator during inhalation and a non-positive operator during exhalation.

We therefore enforce a global non-positivity constraint on the
reconstructed conductivity changes during inhalation, and a global
non-negativity constraint during exhalation, which is named as global monotonicity-based method (GMM). Moreover, on each pixel, we
derive a local lower bound during inhalation (and a local upper bound
during exhalation) using a sensitivity-based variant of the linearized
monotonicity method for inclusion detection \cite{Harrach2013,Tamburrino2006,Tamburrino2002}, which is named as local monotonicity-based method (LMM).  Enforcing the monotonicity constraints in the image reconstruction algorithm can compensate for the inherent ill-conditioned nature of EIT \cite{Ammari2009}.  The effectiveness of the proposed GMM is validated by numerical simulations, phantom experiments and human experiments while LMM is tested only by numerical simulations and phantom experiments.
\section{Methods}
\subsection{Mathematical model}

Let an imaging object occupy a two- or three-dimensional region $\Om \subset \R^n$ $(n=2,3)$. We denote the time-varying conductivity at  position $x$ and  time $t$  by $\sigma^t(x)$. {  It is assumed that $\sigma^{t}(x)\in L^{\infty}_{+}(\Om)$ and $\f{\partial }{\partial t}\sigma^{t}(x)\in L^{\infty}(\Om)$, where $L^{\infty}_{+}(\Om)$ denotes the subspace of $L^{\infty}(\Om)$, functions with positive essential infima.} In $E$-channel EIT system as shown in Fig.\ \ref{Fig-model-data} (a), we attach $E$ electrodes $\mE_1, \mE_2, \cdots, \mE_E$ on its boundary $\p\Om$ to inject $E$ different currents ($E-1$ linearly independent) using orderly chosen pairs of electrodes.  For the ease of explanation, we use the adjacent pair of electrodes $\mE_j$ and $\mE_{j+1}$ to inject $j$-th current. Here and in the following, we use the convention that $\mE_{E+1}=\mE_1$. Then, the distribution of the voltage subject to the $j$-th injection current, denoted by $u_t^j$, is governed by the following  complete electrode model (CEM) \cite{Somersalo1992,Vauhkonen1999}
\begin{eqnarray}\label{NeumannBVP}
&\na \cdot\left( \sigma^t\na u_t^j\right)=0 \qquad\mbox{ in } \Omega,\label{NeumannBVP-1}\\
&\sigma^t\f{\p u_t^j}{\p \n} =0\qquad\mbox{ on }\,\,\p\Om \setminus \bigcup_{k=1}^E \mE_k,\label{NeumannBVP-2}\\
&\left.\left(u_t^j+z_{j,k}\sigma^t\f{\p u_t^j}{\p \n}\right)\right|_{\mE_k}=U^{j,k}(t), \qquad k=1,\ldots,E\label{NeumannBVP-3}\\
&\int_{{\mE}_{k}} \sigma^t\f{\p u_t^j}{\p \n} \,ds  =0\,\qquad \mbox{if}\,\, k\in\{1, \ldots, E\}\backslash\{j, j+1\},\label{NeumannBVP-4}\\
 &\int_{{\mE}_{j}} \sigma^t\f{\p u_t^j}{\p \n} \,ds  =I=-\int_{{\mE}_{j+1}} \sigma^t\f{\p u_t^j}{\p \n} \,ds,\label{NeumannBVP-5}
\end{eqnarray}
where $\n$ is the outward unit normal vector on $\p\Om$ , $U^{j,k}$ is the voltage on $\mE_k$, and $z_{j,k}$ is the contact impedance of the $k$-th electrode $\mathcal{E}_k$.
The magnitude of the current driven through the $j$-th and the $(j+1)$-th electrode is assumed to be normalized to $I=1$.  The solution $u_t^j$ is unique up to constant factors. Let's denote $u_0^j$ as the reference potential, which is the solution of
(\ref{NeumannBVP}) with the  conductivity distribution $\sigma^t$ being replaced by the constant 1.

Assuming that the boundary voltages between all adjacent pairs of electrodes are measured,
the $k$-th boundary voltage measured between $\mE_k$ and $\mE_{k+1}$ subject to the $j$-th injection current is the time-varying function
\begin{equation}
V^{j,k}(t)= U^{j,k}(t)-U^{j,k+1}(t),
\end{equation}
where $k, j=1,2,\ldots, E$. {  Please note that $U^{j,E+1}(t)=U^{j,1}(t)$ and $V^{j,k}(t)=V^{k,j}(t)$}.
Thus, we collect $E^2$ number of time-varying boundary data which can be expressed as the following symmetric matrix form
\begin{equation}\label{Eq:TimeData}
{\mathbb V}(t) =\left[\begin{array}{ccc}
V^{1,1}(t) &\cdots &V^{1,E}(t)\\
 V^{2,1}(t) & \cdots& V^{2,E}(t)\\
 \vdots& \ddots&\vdots \\
 V^{E,1}(t)& \cdots& V^{E,E}(t)
 \end{array}\right].
\end{equation}
The inverse problem of lung EIT for monitoring the ventilation  is to visualize the time varying distribution of $\sigma^t$ in the lung regions from the time-varying  data ${\mathbb V}(t)$.

\subsection{Boundary data separation}\label{subsection-data-separation}
It is reasonable to assume  $z_{j,k}\sigma^t\f{\p u_t^j}{\p \n}\approx0$ on the voltage sensing electrodes in  (\ref{NeumannBVP-1})-(\ref{NeumannBVP-5}), so that we have $u_t^j|_{\mE_k}=U^{j,k}(t)$ for  $|k-j|>1$.
Therefore,  for $|k-j|>1$,  the conductivity is related to the measured signals by the following identities
\begin{eqnarray}
 V^{j,k}(t)&=\int_{\Om} \sigma^t \na u_t^j\cdot\na u_t^k dx,\label{Eq:T-data_standard}\\
\f{d}{dt} V^{j,k}(t)&=-\int_{\Om} \f{\p\sigma^t}{\p t} \na u_t^j\cdot\na u_t^k dx.\label{Eq:T-data_standard_diff}
\end{eqnarray}
{  Note that if we further assume the electrodes are perfectly conductive, i.e., $z_{j,k}=0$ for all $j, k= 1,2,\cdots,E$, the complete electrode model in (\ref{NeumannBVP-1})-(\ref{NeumannBVP-5}) becomes the so-called shunt model \cite{Cheney1999}.  Then (\ref{Eq:T-data_standard}) and (\ref{Eq:T-data_standard_diff}) hold for $|k-j|\leq1$ as well, which are proven in the Appendix A.1. Throughout this paper, the theory development are based on the shunt model.}

However, the unknown contact impedances of current-driven electrodes are not ignorable in real application. The measurements on the current-driven electrodes are affected by the unknown contact impedances. Therefore, the identities (\ref{Eq:T-data_standard}) and (\ref{Eq:T-data_standard_diff}) fail for $|k-j|\leq1$.

The potentials $u_t^j$ and $u_t^k$ are related to the unknown
conductivity $\sigma^t$ via (\ref{NeumannBVP}). { Standard linearized reconstruction methods (LM) \cite{Seo2012book} for lung EIT replace $u_t^j$ and $u_t^k$ on the right hand
side of (\ref{Eq:T-data_standard_diff}) by the reference potentials $u_{0}^{j}$ and $u_{0}^{k}$ and solve the resulting linear equation to determine
$\f{\p\sigma^t}{\p t}$ from $\f{d}{dt} V^{j,k}(t)$ by relation
\begin{equation}\label{eq-linearization}\f{d}{dt} V^{j,k}(t)\approx -\int_{\Om} \f{\p\sigma^t}{\p t} \na u_0^j\cdot\na u_0^k dx.\end{equation}}

\begin{figure*}[ht]
\centering
\setlength\fboxsep{0pt}
\setlength\fboxrule{0.25pt}
  \begin{center}
\begin{tabular}{ccc}
\includegraphics[width=4.5cm]{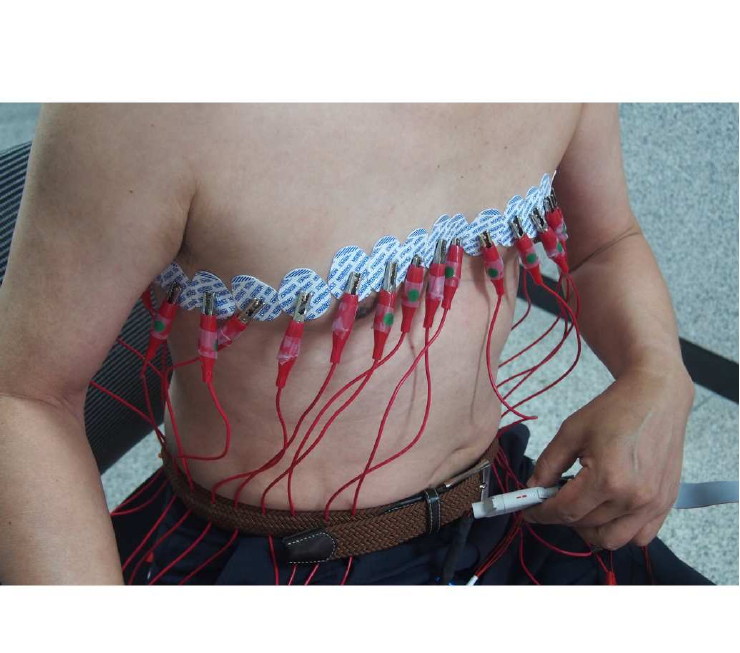}&
\includegraphics[width=4.5cm]{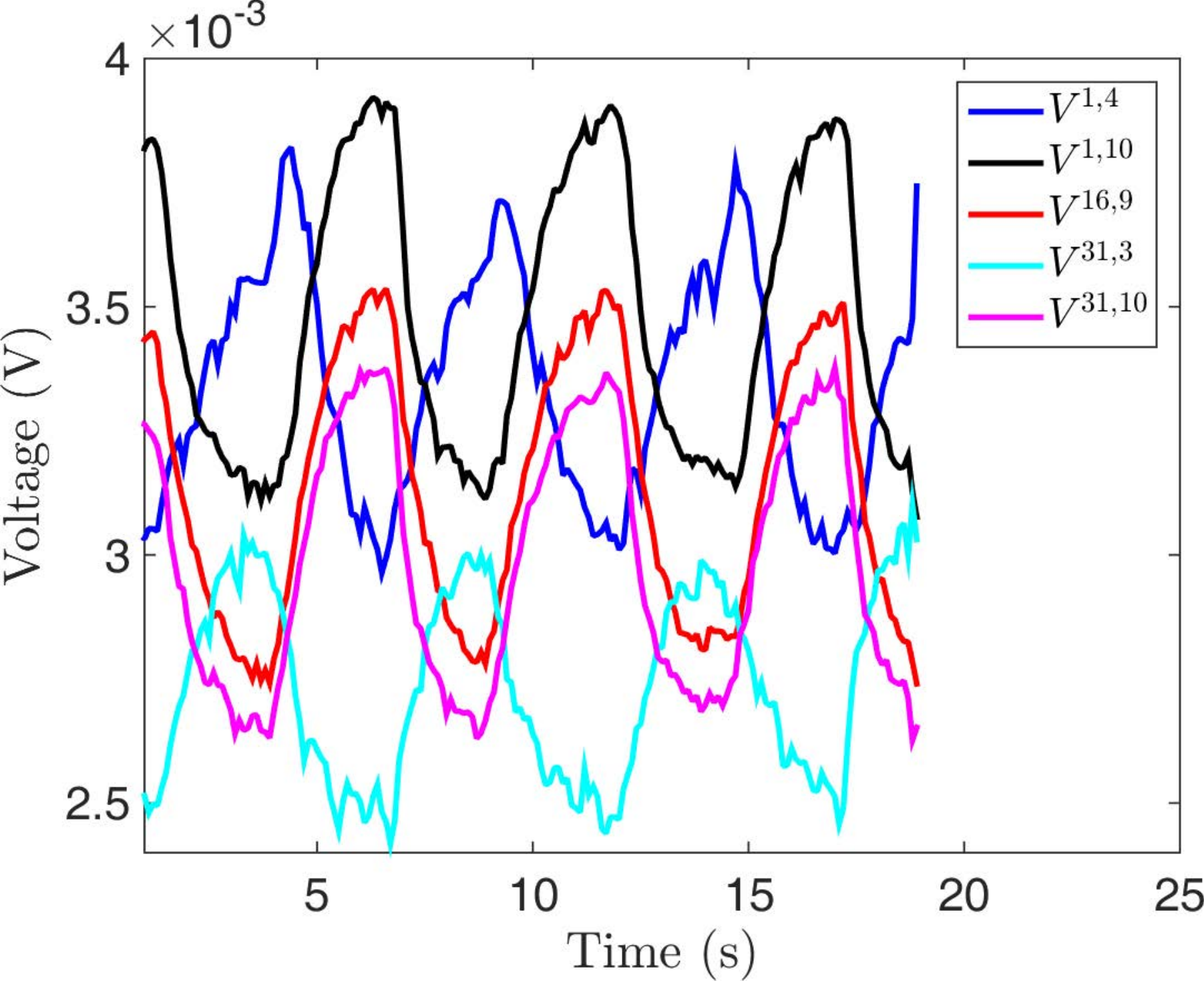}&
\includegraphics[width=4.5cm]{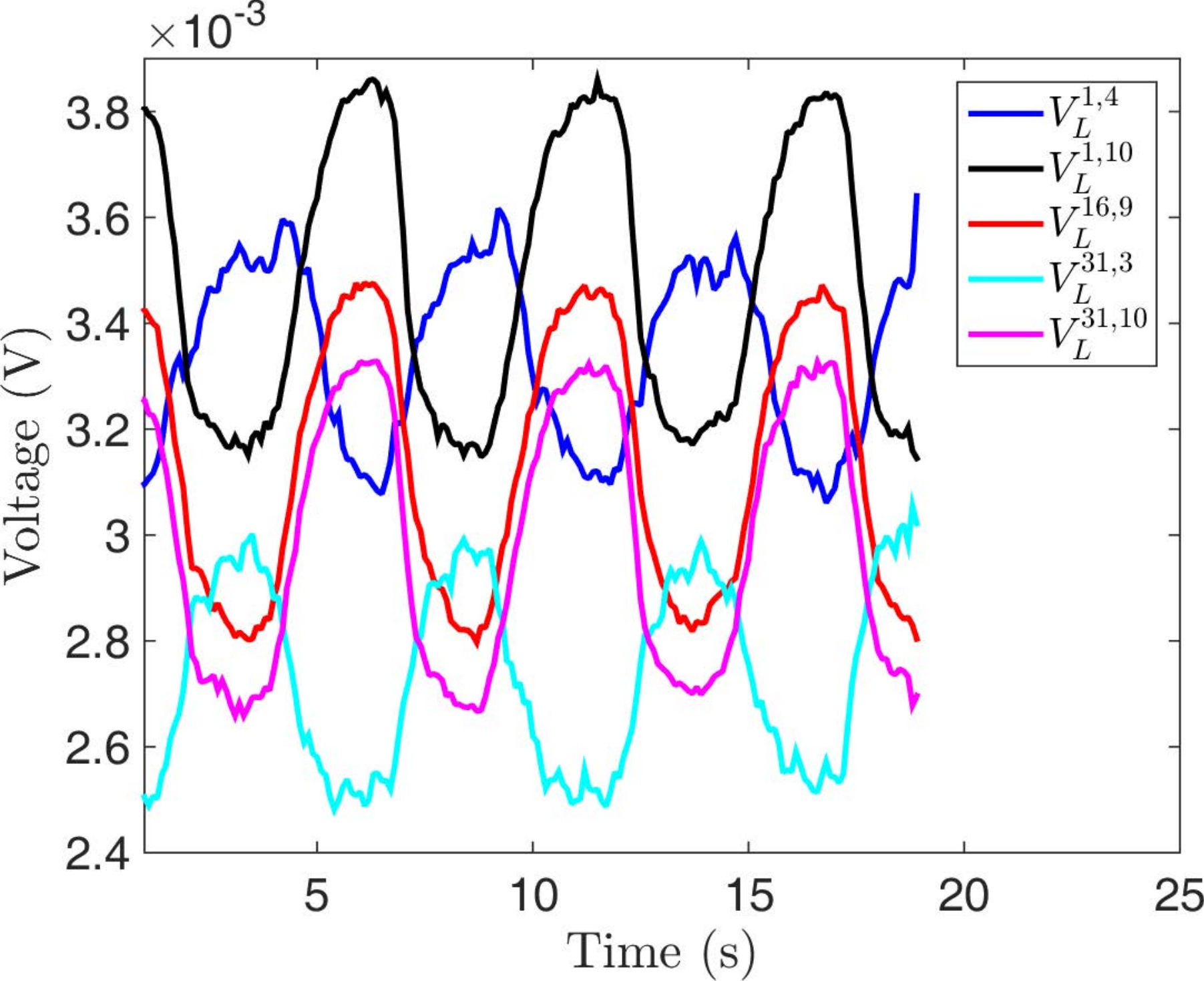}\\
(a)& (b)&(c)\\
\includegraphics[width=4.5cm]{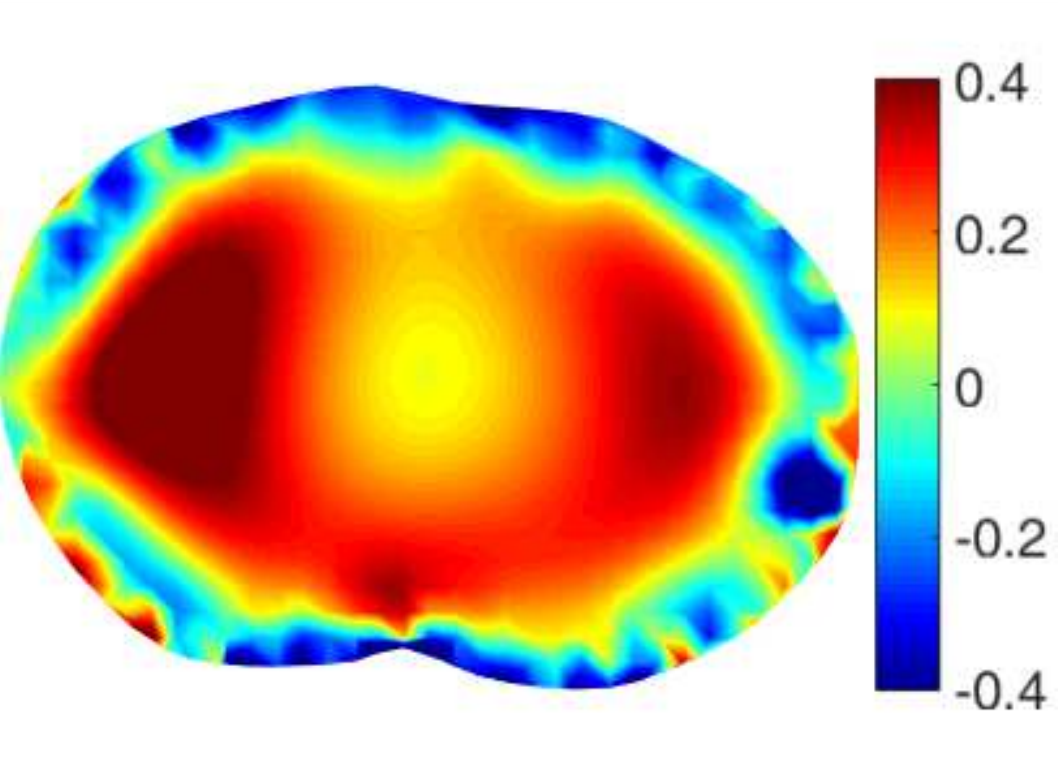}&
\includegraphics[width=4.5cm]{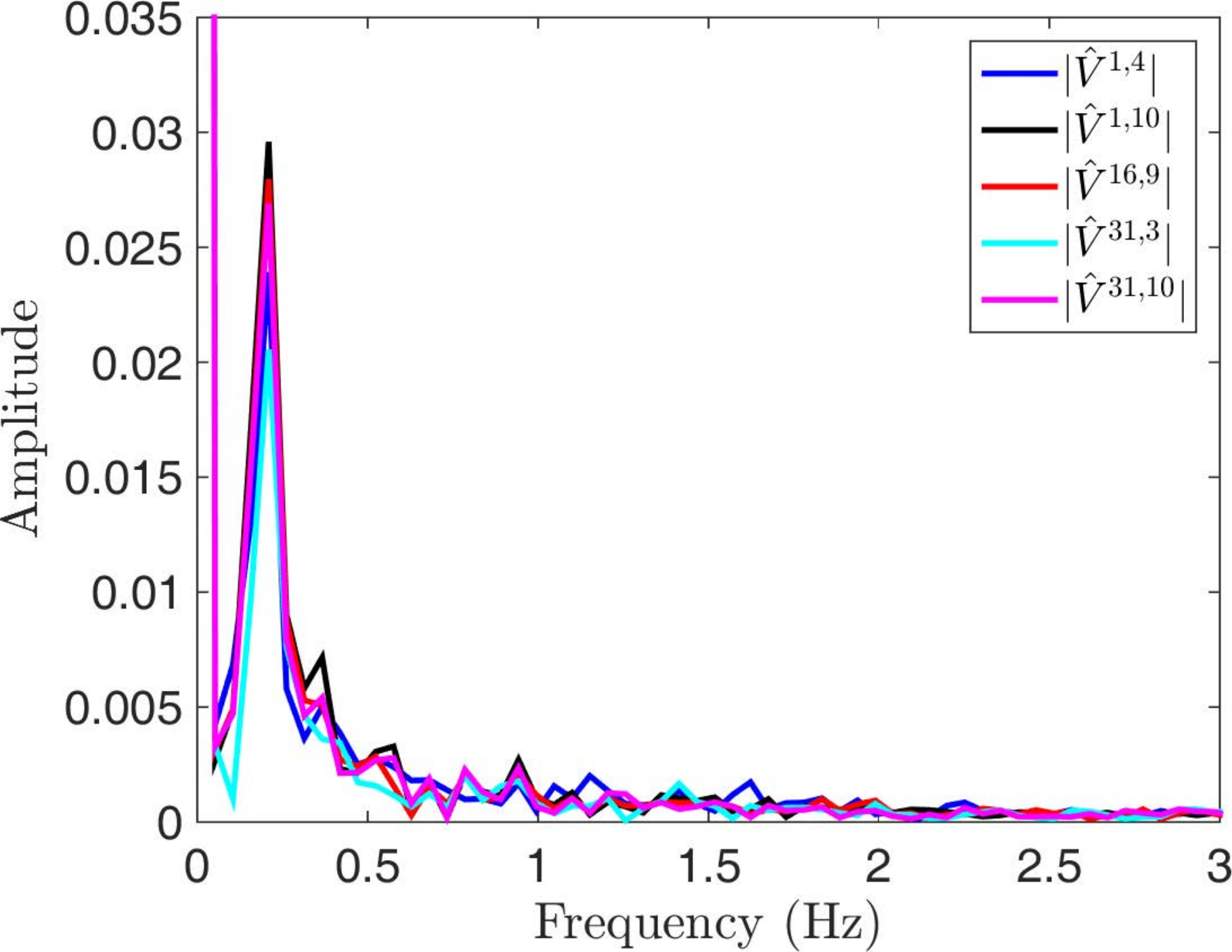}&
\includegraphics[width=4.5cm]{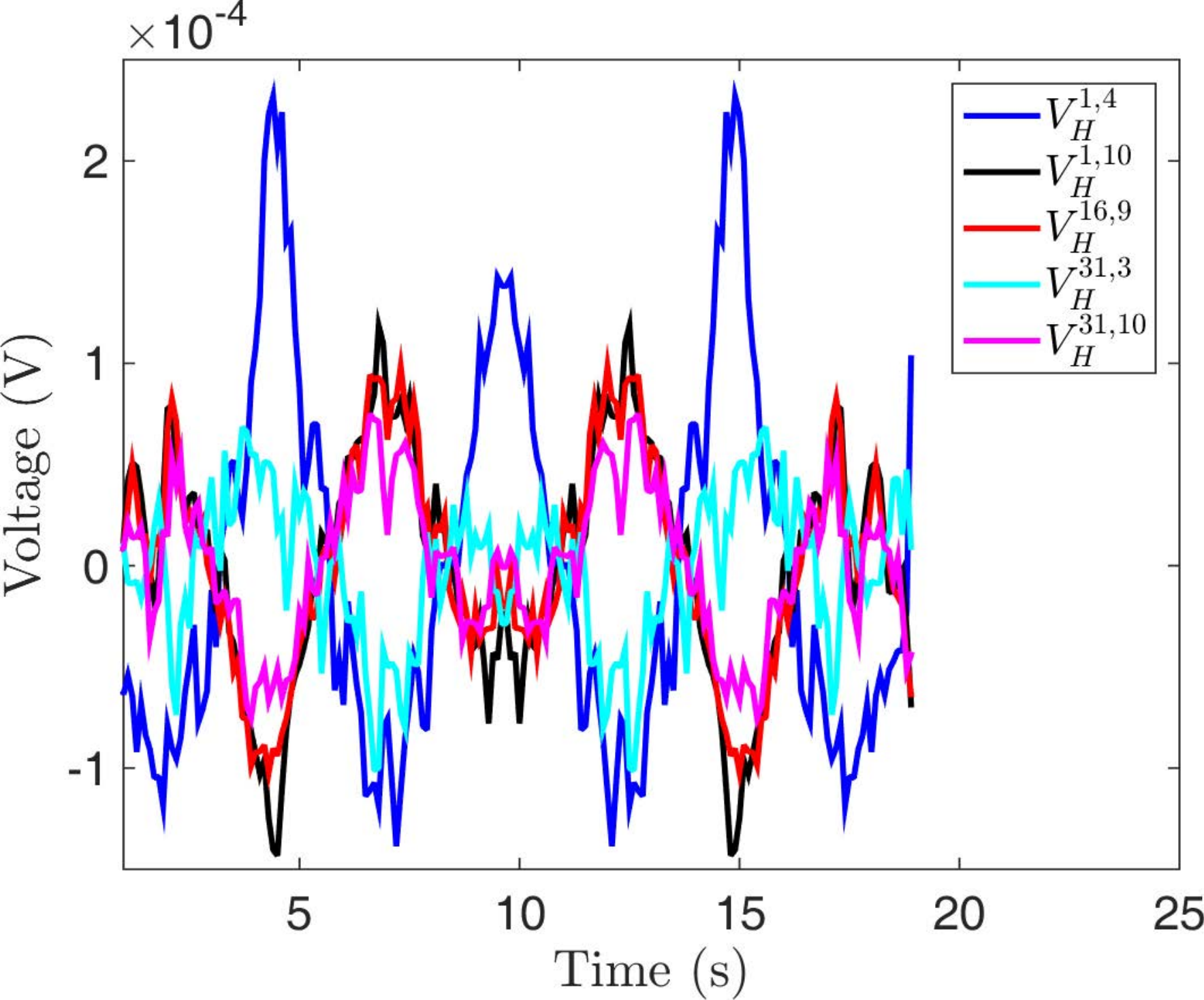}\\
(d)&(e)&(f)
\end{tabular}
\end{center}
\caption{EIT data provided by 32-channel Swisstom EIT system with 4 skipped injection pattern. In this special pattern, $V^{j,k}$ denotes the voltage difference between $\mE_{k}$ and $\mE_{k+5}$ subject to the injection current using the pair $\mE_j$ and $\mE_{j+5}$.  (a) electrodes configuration; (b) measured boundary data $V^{j,k}$; (c) low frequency component $V_L^{j,k}$; (d) reconstructed conductivity image using the standard LM; (e) Fast Fourier transform $\hat V^{j,k}$;  (f) high frequency component  $V_H^{j,k}$.}
\label{Fig-model-data}
\end{figure*}

In this work, we will propose two monotonicity-based improvements to this standard linearized reconstruction method in (\ref{eq-linearization}).
The first improvement is based on the following data separation approach. We decompose the time-derivative of the conductivity $\f{\p}{\p t} \sigma^t(x)$ at position $x$ into two parts
\begin{equation}\label{Eq:conductivity-decom}
\f{\p}{\p t}\sigma^t(x)=\f{\p}{\p t}\sigma_L^t(x)+\f{\p}{\p t}\sigma_H^t(x)\quad\,\,x\in\Om,
\end{equation}
where $\f{\p}{\p t}\sigma_L^t$ and $\f{\p}{\p t}\sigma_H^t$ are conductivity change induced by the  lung ventilation, cardiac activities and other factors, respectively.

For the application of lung monitoring, we are interested in recovering the ventilation-induced conductivity change $\f{\p\sigma_L^t}{\p t}$. { Hence, it follows from (\ref{Eq:T-data_standard})-(\ref{Eq:conductivity-decom}) that we
reasonably assume the existence of a ventilation-induced signal}
\begin{equation}\label{000}
V^{j,k}_L(t)= \int_{\Om} \sigma_L^t \na u_{t}^j\cdot\na u_{t}^k dx,
\end{equation}
and aim to extract
\begin{equation}\label{111}
\f{d}{dt} V^{j,k}_L(t)= -\int_{\Om}\f{\p\sigma_L^t}{\p t}\na u_{t}^j\cdot\na u_{t}^k dx,
\end{equation}
from $\f{d}{dt} V^{j,k}(t)$.  Similar as (\ref{Eq:TimeData}), the ventilation-induced data can be written as an $E\times E$ matrix
\begin{equation}\label{Eq:LungData}
{\Bbb V}_L(t) =\left[\begin{array}{ccc}
V^{1,1}_L(t) &\cdots &V^{1,E}_L(t)\\
 V^{2,1}_L(t) & \cdots& V^{2,E}_L(t)\\
 \vdots& \ddots&\vdots \\
 V^{E,1}_L(t)& \cdots& V^{E,E}_L(t)
 \end{array}\right].
\end{equation}

{ Due to the clear distinction of frequencies for pulmonary (~ $12~{min}^{-1}$) and cardiac (~$60-80~{min}^{-1}$) activities, the ventilation-induced signals $V^{j,k}_L(t)$ can be obtained by using Fast Fourier transform and applying band-pass filter to the measured data $V^{j,k}(t)$ with a selected frequency range \cite{Grant2011}, see Fig. \ref{Fig-data-separation} for numerical validation. The cutoff frequencies $8~min^{-1}$ and $20 ~min^{-1}$ were used for the band-pass filter to obtain $V^{j,k}_L(t)$, which are around the breath rate $12~ {min}^{-1} $ for healthy adult \cite{Frerichs2009,Grant2011}}. Fig. \ref{Fig-model-data} shows a human experiment.  We see that the separated ventilation-induced signals in Fig. \ref{Fig-model-data} (c)  is periodic and monotonically changing in the period of inhalation and exhalation.  Fig. \ref{Fig-model-data} (f) shows the high frequency component $V_H^{j,k}$ which is corresponding to the cardiac activities, measurement noise and other factors. The cardiac signal peak is not obviously shown in Fig. \ref{Fig-model-data} (e) because the cardiac signal is significantly smaller in amplitude compare with the ventilation-induced signal.

\subsection{Monotonicity constraints}
After extracting the ventilation-induced signals $V^{j,k}_L(t)$ from the measured signals using a band-pass filter, we can reconstruct
the ventilation-induced conductivity change $\f{\p\sigma_L^t}{\p t}$ by approximating the potentials in (\ref{111}) by reference potentials and solving the
resulting linear equation. We improve this strategy with the following monotonicity-based constraints. { By considering the air inhalation and exhalation during breath, it is reasonable to assume that $\sigma^{t}_{L}$ is monotone with respect to $t$ in the sense that the ventilation-induced conductivity is either increasing everywhere during exhalation,
\begin{equation}\label{Eq:sigma_increasing}
\f{\p}{\p t}\sigma_L^t(x)\geq 0 \quad\hbox{for all}~ x\in\Om,
\end{equation}
or decreasing everywhere during inhalation,
\begin{equation}\label{Eq:sigma_decreasing}
\f{\p}{\p t}\sigma_L^t(x)<0 \quad\hbox{for all}~ x\in\Om.
\end{equation}}
 { We should mention that the above monotonicity hypothesis may not always be entirely true for all the cases due to the inhomogeneity of lung tissue and the change of outer shape of the domain.} However, the monotonicity assumption makes sense when considering pressure changes during inspiration and expiration. During inhalation, air flows into the lungs as a result of decrease in the pressure inside the lungs. As a consequence, the overall effective conductivity in the lungs decreases at inhalation \cite{Chiras2010,Nopp1997}. Similarly, the overall effective conductivity in the lungs increases at exhalation.  The use of the monotonicity constraint as a temporal regularization allows to deal with the ill-posed nature of EIT and improve the robustness in reconstruction.

From (\ref{111}), we have
\begin{eqnarray}\label{Eq:LinearEq}
{\bf a}^T\f{d}{dt}{\Bbb V}_L(t){\bf a}
\q= -\int_{\Om}\f{\p\sigma_L^t}{\p t}\na \left(\sum_{j=1}^Ea_j u_{t}^j\right)\cdot\na \left(\sum_{k=1}^Ea_k u_{t}^k\right)dx\label{111-2}
\end{eqnarray}
for all vectors ${\bf a}=(a_1,\ldots,a_E)^T\in \R^E$. For a symmetric matrix $A$, we write $A\succeq 0$ if $A$ is semi-positive definite, that is, ${\bf a}^T A {\bf a}\ge 0$ for all  vectors ${\bf a}$. We write $A\preceq 0$ if $A$ is semi-negative definite.
From (\ref{Eq:LinearEq}), we have the following monotonicity relations
\begin{eqnarray}
&&\inf_{x\in\Om}\f{\p\sigma_L^t}{\p t}(x)\geq0~~\Rightarrow~~ \f{d }{d t}{\Bbb V}_L\preceq0,
\label{Eq:Mono-relation-1}\\
&&\sup_{x\in\Om}\f{\p\sigma_L^t}{\p t}(x)\leq0~~\Rightarrow~~ \f{d }{d t}{\Bbb V}_L\succeq 0. \label{Eq:Mono-relation-2}
\end{eqnarray}
Here, the right sides of  (\ref{Eq:Mono-relation-1}) and (\ref{Eq:Mono-relation-2}) should be understood in the sense of matrix definiteness.

The proposed methods evaluate whether the conductivity of the lungs is in the increasing or decreasing period and reconstruct the lung ventilation-induced conductivity change $\f{\p\sigma_L^t}{\p t}$ using the additional global monotonicity-based constraint (\ref{Eq:sigma_increasing}), resp., (\ref{Eq:sigma_decreasing}). Given the monotonicity assumption in (\ref{Eq:sigma_increasing}) and (\ref{Eq:sigma_decreasing}), it has been shown in \cite{Harrach2013} that testing the matrix definiteness on the right hand sides of (\ref{Eq:Mono-relation-1}) and (\ref{Eq:Mono-relation-2}) can give the outer shape of a conductivity change for continuous boundary measurements. For electrode measurements one can expect to obtain an upper bound of the conductivity
change, cf., \cite{HLU15, HM16a, HM16b, BHHM17,Gar17, GS17a, GS17b} for recent works on this monotonicity-based approach in electrical impedance tomography.

\subsection{EIT data sources consistency} \label{Sub-source}
For the monotonicity-based spatiotemporal EIT imaging, it is necessary to determine the  conductivity monotone increase and monotone decrease period.  According to (\ref{Eq:Mono-relation-1}) and (\ref{Eq:Mono-relation-2}), the conductivity increasing period can be determined by evaluating the definiteness of $\f{d }{d t}{\Bbb V}_L$. However, due to the unknown contact impedances of the current driven electrodes, the tridiagonal elements of ${\Bbb V}_L$ are not reliable for determining the period.
 Fortunately,  we made an interesting  observation about EIT data sources consistency through several human experiments. This  allows us to determine the monotone increase and monotone decrease period of conductivity.

\begin{figure}[ht!]
\centering
\begin{tabular}{p{7cm}p{7cm}}
\includegraphics[width=6cm]{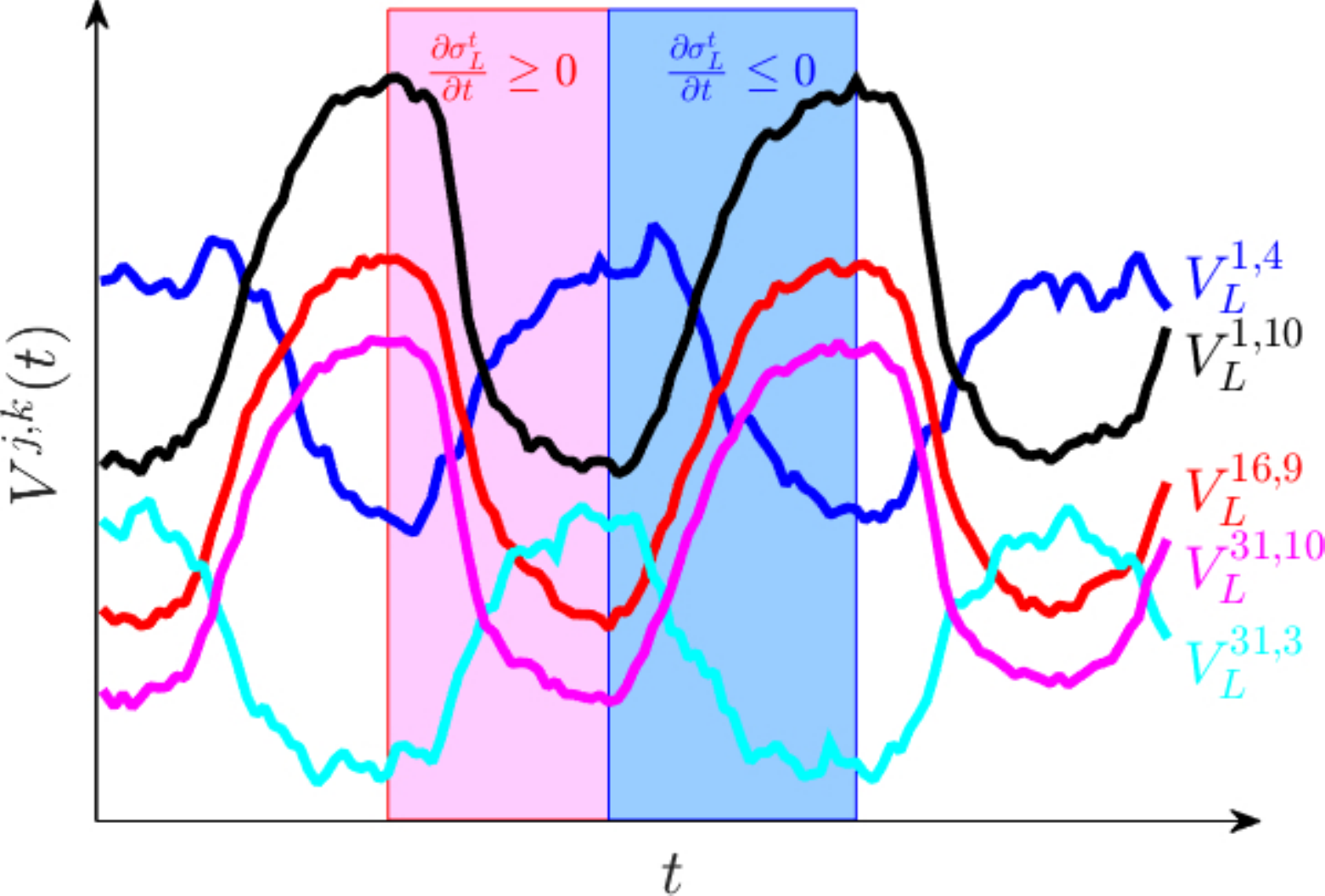}&
\includegraphics[width=6cm]{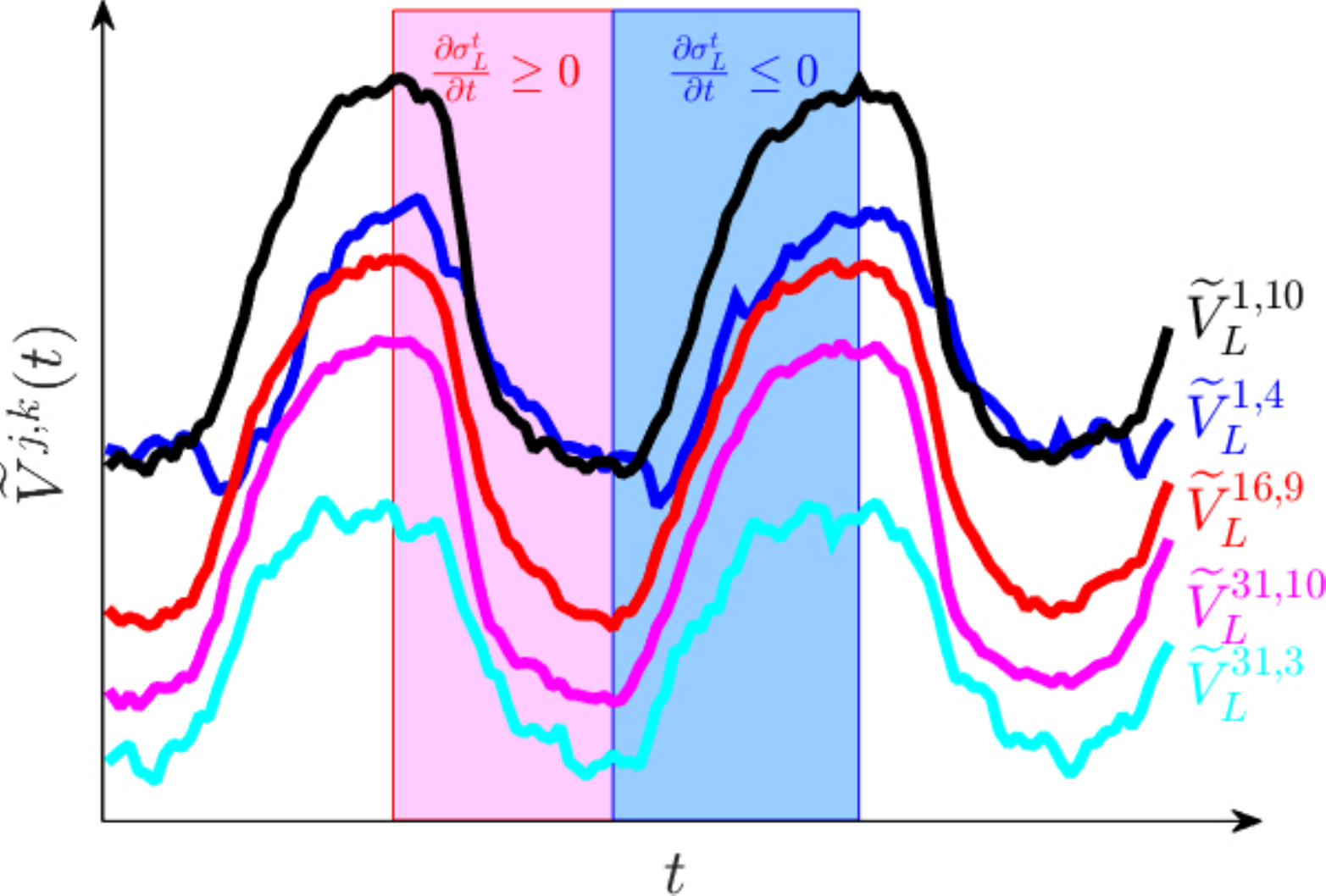}\\
\qquad\qquad\quad(a)&\qquad\qquad\quad(b)
\end{tabular}
\caption{EIT data obtained from human experiment using 32-channel Swisstom EIT system with 4 skipped injection pattern. (a) band-pass filtered data ${ V}_L^{j,k}$; (b) transformed data ${\widetilde V}_L^{j,k}$ by formula (\ref{Eq:observation}).}
\label{Fig-mono-data}
\end{figure}

This observation is obtained by a simple modification of the data $V_L^{j,k}(t)$ based on the equations (\ref{000}), (\ref{111}) and the monotonicity assumptions (\ref{Eq:sigma_increasing}) and (\ref{Eq:sigma_decreasing})
\begin{eqnarray}
{\widetilde V_L^{j,k}}(t):&=\f{V_L^{j,k}(t)}{{sgn}\Big(\int_{\Om}\na u^j_0\cdot\na u^k_0dx\Big)}\label{Eq:observation}\\
&\quad+\Big(1-{sgn}\Big(\int_{\Om}\na u^j_0\cdot\na u^k_0dx\Big)\Big){ave}(V_L^{j,k}),\nonumber
\end{eqnarray}
where $sgn$ is the sign function and ${ave}(V_L^{j,k})$  indicates the average of  $V_L^{j,k}$ over one  period of time.
Fig. \ref{Fig-mono-data} shows $V_L^{j,k}(t)$ and ${\widetilde V_L^{j,k}}(t)$, which are obtained from human experiment.
As shown in Fig. \ref{Fig-mono-data} (b), most of ${\widetilde V_L^{j,k}}$ are of the same pattern in terms of increasing and decreasing period.

{ Several human experiments (seven adults in our lab) and numerical experiments showed that the time change  of ${\widetilde V}_L^{j,k}$ (except for a few pair of ($j, k$)) has the same pattern as $\sigma_L^t$, as shown in Fig. \ref{Fig-mono-data}  (b), which could help us determine the monotonicity period by observing the pattern of the modified data. It is likely that the few pair of inconsistent data in ${\widetilde V}_L^{j,k}$ is related to $\f{\int_{\Om}\na u_0^j\cdot \nabla u_0^kdx}{\int_{\Om}|\na u_0^j\cdot \nabla u_0^k|dx}\leq c$ for a constant $c<0$, meaning  that the sensitivity $\na u_0^j\cdot \nabla u_0^k$ is negative and overwhelms positive sensitivity by some amount.
In Fig. \ref{Fig-mono-data}, the blue rectangular area shows $\f{d}{dt}(\sum_{j,k}\widetilde V_{L}^{j,k}(t))\geq0$ which indicates the period of decrease in conductivity ($\f{\p\sigma_L^t}{\p t}\leq0$). On the other hand, the pink rectangular area shows $\f{d}{dt}(\sum_{j,k}\widetilde V_{L}^{j,k}(t))\leq 0$ which indicates the period of increase in conductivity ($\f{\p\sigma_L^t}{\p t}\geq0$).}

  The consistency between modified data $\widetilde V_L^{j,k}$ and conductivity $\sigma_L^t$ is of great importance to determine  the period of increase and decrease in conductivity for imposing temporal regularization constraints.  Future studies will be needed to provide a rigorous proof of  this observation.

\subsection{Imaging with global monotonicity-based constraints}\label{subsection-global-recon}
Discretizing the {  reference imaging domain $\bar\Om$ into $\bar\Om=\cup_{p=1}^{N}T_p$ }where $T_p$ is the $p$-th pixel, the conductivity distribution $\sigma_L^t$ in domain $\bar\Om$ can be expressed as $N\times 1$ vector ${\boldsymbol\sigma}_L^t$ (bold symbol). For a vector ${\bf b}=(b_1,\ldots, b_{N})^T$, we write ${\bf b}\geq 0$, resp., ${\bf b}\leq 0$ if  all components of the vector are non-negative, resp., non-positive. From the monotonicity assumption
(\ref{Eq:sigma_increasing}), resp., (\ref{Eq:sigma_decreasing}), the time differential $\f{\p{\boldsymbol\sigma}_L^t}{\p t}$ satisfies the following uniform positivity (or negativity) property
\begin{eqnarray}
\label{Eq:monotonicity-L2}
\mbox{either}~~\f{\p{\boldsymbol\sigma}_L^t}{\p t}\geq0   ~~\mbox{or}~~\f{\p{\boldsymbol\sigma}_L^t}{\p t}<0.
\end{eqnarray}

This paper uses this monotonicity constraint (\ref{Eq:monotonicity-L2}) to solve the ill-conditioned linear system (called linearized EIT system)
that arises from approximating the potentials in (\ref{111}) by reference potentials
\begin{eqnarray}\label{Eq:LinearSystem}
\f{d}{d t}{\bf V}_L~=~\mathbb{S}\,\,\f{\p {\boldsymbol\sigma}_L^t}{\p t},
\end{eqnarray}
where $\mathbb{S}$ is the $E^2\times N$ sensitivity matrix whose $((j-1)*E+k, p)$ element is $S_p^{j,k} = -\int_{T_p}\na u_{0}^j\cdot\na u_{0}^kdx$, and ${\bf V}_L(t)$ is the column concatenated data set of $\Bbb V_L(t)$ given by
\begin{equation}
\label{Vdata}
{\bf V}_L=(V^{1,1}_L, \ldots, V^{1,E}_L, \ldots,V^{E,1}_L, \ldots, V^{E,E}_L)^T.
\end{equation}

We obtain the following global monotonicity-based constrained reconstruction method (GMM)
\begin{itemize}
\item In the conductivity increasing period
\begin{eqnarray}\label{Eq:min-increase}
&\mbox{min} ~\left\| \mathbb{S}\,\f{\p {\boldsymbol\sigma}_L^t}{\p t}-\f{d}{d t}{\bf V}_L\right\|_2^2+\lambda\|\f{\p {\boldsymbol\sigma}_L^t}{\p t}\|_2^2 \quad\mbox{subject to}~~ \f{\p{\boldsymbol\sigma}_L^t}{\p t}\geq0,
\end{eqnarray}
\item  In the  conductivity decreasing period
\begin{eqnarray}\label{Eq:min-decrease}
&\mbox{min} ~\left\| \mathbb{S}\,\f{\p {\boldsymbol\sigma}_L^t}{\p t}-\f{d}{d t}{\bf V}_L\right\|_2^2+\lambda\|\f{\p {\boldsymbol\sigma}_L^t}{\p t}\|_2^2 \quad\mbox{subject to}~~ \f{\p{\boldsymbol\sigma}_L^t}{\p t}\leq0,
\end{eqnarray}
\end{itemize}
where $\lambda\geq0$ is the regularization parameter.

In order to implement the minimization problems of (\ref{Eq:min-increase}) and (\ref{Eq:min-decrease}), we discretize time $t$ as
\[
t_1<\cdots<t_{n-1}<t_{n+1}<\cdots,
\]
and reconstruct the time difference ${\boldsymbol\sigma}_L^{t_n}-{\boldsymbol\sigma}_L^{t_{n-1}}$ by solving the following constrained minimization problems
\begin{itemize}
\item In the  conductivity increasing  period
\begin{eqnarray}\label{Eq:min-increase-1}
&\mbox{min} ~\left\| \mathbb{S}\, ({\boldsymbol\sigma}_L^{t_n}-{\boldsymbol\sigma}_L^{t_{n-1}})-({\bf V}_L^{t_n}-{\bf V}_L^{t_{n-1}})\right\|_2^2+\lambda\|{\boldsymbol\sigma}_L^{t_n}-{\boldsymbol\sigma}_L^{t_{n-1}}\|_2^2\\
&\mbox{subject to}~~ {\boldsymbol\sigma}_L^{t_n}-{\boldsymbol\sigma}_L^{t_{n-1}}\geq0,
\end{eqnarray}
\item In the conductivity decreasing  period
\begin{eqnarray}\label{Eq:min-decrease-1}
&\mbox{min} ~\left\| \mathbb{S}\, ({\boldsymbol\sigma}_L^{t_n}-{\boldsymbol\sigma}_L^{t_{n-1}})-({\bf V}_L^{t_n}-{\bf V}_L^{t_{n-1}})\right\|_2^2+\lambda\|{\boldsymbol\sigma}_L^{t_n}-{\boldsymbol\sigma}_L^{t_{n-1}}\|_2^2\\
&\mbox{subject to}~~ {\boldsymbol\sigma}_L^{t_n}-{\boldsymbol\sigma}_L^{t_{n-1}}\leq0.
\end{eqnarray}
\end{itemize}

\subsection{Additional local monotonicity-based constraints}\label{Subsetion-local}

From subsection \ref{subsection-data-separation} to subsection \ref{subsection-global-recon}, we used a data splitting approach and a monotonicity argument to obtain a global lower (upper)  bound
for the conductivity change in the conductivity increasing (decreasing)  period. In this subsection, we use a refined monotonicity argument
to obtain also an upper bound for the conductivity change for the  conductivity increasing period(and, analogously, a lower bound for the  conductivity decreasing period). The following approach is a new combination of the standard linearized
 reconstruction method with a sensitivity based variant of the
 monotonicity method developed in \cite{Harrach2013}. This is
 motivated by the recent paper \cite{choi2014regularizing} (see also
 \cite{harrach2010factorization}) which uses a sensitivity based variant
 of the factorization method (see
 \cite{harrach2013recent,hanke2011sampling,kirsch2008factorization}) to
 regularize the standard linearized reconstruction method.

Given the $p$-th pixel $T_p\subset\bar\Om$, define the following $E\times E$ matrix
\begin{equation*}
S_p^t=
-\left[\begin{array}{ccc}\int_{T_p}\na u_{t}^1\cdot\na u_{t}^1dx&\cdots&\int_{T_p}\na u_{t}^1\cdot\na u_{t}^Edx\\
\vdots&\ddots&\vdots\\
\int_{T_p}\na u_{t}^E\cdot\na u_{t}^1dx&\cdots&\int_{T_p}\na u_{t}^E\cdot\na u_{t}^Edx\end{array}\right]
\end{equation*}
and note that, for ${\bf a}=(a_1,\ldots, a_E)^T\in\mathbb{R}^E$, we have
\begin{equation}\label{Eq:S_p_quad_form}
-{\bf a}^T S_p^t {\bf a}=\int_{T_p} \left| \sum_{j=1}^Ea_j \na u_{t}^j \right |^2 dx.
\end{equation}

We will use the following quantitative version of the monotonicity relations (\ref{Eq:Mono-relation-1}) and (\ref{Eq:Mono-relation-2}). For ${\bf a}=(a_1, \ldots, a_E)^T\in\mathbb{R}^E$,
\begin{eqnarray}
\hspace{-1cm}\nonumber\int_{\bar\Om} (\sigma_L^{t_{n-1}}-\sigma_L^{t_{n}}) \left| \sum_{j=1}^Ea_j \na u_{t_{n}}^j \right |^2 dx&\geq {\bf a}^T({\Bbb V}_L(t_n)-{\Bbb V}_L(t_{n-1})){\bf a}\\
 &\geq \int_{\bar\Om} (\sigma_L^{t_{n-1}}-\sigma_L^{t_{n}}) \left| \sum_{j=1}^Ea_j \na u_{t_{n-1}}^j \right |^2 dx,\label{Eq:estimate1}
\end{eqnarray}
which is proven in the Appendix A.2.

Consider the  conductivity increasing period, $\sigma_L^{t_{n-1}}\leq \sigma_L^{t_{n}}$.
If $\alpha\geq0$ fulfills
\[
-\alpha \chi_{T_p} \geq \sigma_L^{t_{n-1}} - \sigma_L^{t_{n}},
\]
then, by (\ref{Eq:S_p_quad_form}) and (\ref{Eq:estimate1}),
\[
\alpha S_p^{t_n} \succeq {\Bbb V}_L(t_n)-{\Bbb V}_L(t_{n-1}),
\]
where $\chi_{T_p}$ is the characteristic function of subset $T_p\subset\bar\Om$.

Under the assumption that the conductivity change is constant on the pixel $T_p$ we hence obtain by contraposition that
\begin{eqnarray}
&&\alpha S_p^{t_n} \not\succeq {\Bbb V}_L(t_n)-{\Bbb V}_L(t_{n-1}) \Rightarrow \ ( \sigma_L^{t_{n}} - \sigma_L^{t_{n-1}})|_{T_p}  \leq \alpha. \label{Eq:evaluate-alpha}
\end{eqnarray}
Hence, for $p$-th pixel we aim to find a smallest possible $\alpha\geq 0$ with $\alpha S_p^{t_n} \not \succeq {\Bbb V}_L(t_n)-{\Bbb V}_L(t_{n-1})$.
Collecting the values for all pixels in a vector $\boldsymbol\alpha$, we arrive at the following local monotonicity-based constrained reconstruction method (LMM).
\begin{itemize}
\item In the  conductivity increasing period
\begin{eqnarray}\label{Eq:min-alpha}
&\mbox{min} ~\left\| \mathbb{S}\, ({\boldsymbol\sigma}_L^{t_n}-{\boldsymbol\sigma}_L^{t_{n-1}})-({\bf V}_L^{t_n}-{\bf V}_L^{t_{n-1}})\right\|_2^2+\lambda\|{\boldsymbol\sigma}_L^{t_n}-{\boldsymbol\sigma}_L^{t_{n-1}}\|_2^2\\
&\mbox{subject to}~~ \boldsymbol\alpha \geq {\boldsymbol\sigma}_L^{t_n}-{\boldsymbol\sigma}_L^{t_{n-1}}\geq0.
\end{eqnarray}
\end{itemize}

Analogously, we obtain for the  conductivity decreasing period, $\sigma_L^{t_{n}}\leq \sigma_L^{t_{n-1}}$, that, for all $\beta\geq 0$,
\begin{eqnarray}
&&-\beta S_p^{t_{n-1}} \not \preceq{\Bbb V}_L(t_n)-{\Bbb V}_L(t_{n-1})\Rightarrow \ ( \sigma_L^{t_{n-1}} - \sigma_L^{t_{n}}  )|_{T_p}  \leq \beta,\label{Eq:evaluate-beta}
\end{eqnarray}
so that we aim to find a smallest possible $\beta\geq 0$ with $-\beta S_p^{t_{n-1}} \not \preceq {\Bbb V}_L(t_n)-{\Bbb V}_L(t_{n-1})$ for $p$-th pixel
and collect the values for all pixels in a vector $\boldsymbol\beta$.
\begin{itemize}
\item In the conductivity decreasing period
\begin{eqnarray}\label{Eq:min-beta}
&\mbox{min} ~\left\| \mathbb{S}\, ({\boldsymbol\sigma}_L^{t_n}-{\boldsymbol\sigma}_L^{t_{n-1}})-({\bf V}_L^{t_n}-{\bf V}_L^{t_{n-1}})\right\|_2^2+\lambda\|{\boldsymbol\sigma}_L^{t_n}-{\boldsymbol\sigma}_L^{t_{n-1}}\|_2^2\\
&\mbox{subject to}~~ -\boldsymbol\beta \leq {\boldsymbol\sigma}_L^{t_n}-{\boldsymbol\sigma}_L^{t_{n-1}}\leq0.
\end{eqnarray}
\end{itemize}

For the implementation of these additional local monotonicity-based constraints, we approximate the potentials $u_t^j$ in the definition
of $S_p^t$  by reference potentials $u_0^j$ as this is done in the above-described linearized reconstruction methods.
Accordingly, $S_p^{t_n}$ and $S_p^{t_{n-1}}$ are approximated by the matrix $S_p$, which is obtained by a rearrangement of the $p$-th column of the sensitivity matrix $\mathbb{S}$ defined in subsection \ref{subsection-global-recon}. The smallest possible values for $\alpha$, resp., $\beta$ on each pixel are
determined by a binary search method.

\section{Algorithm summary}
The constrained minimization problem with the non-negative constraint (\ref{Eq:min-increase-1}) and (\ref{Eq:min-decrease-1}) can be viewed as non-negative least square  problem  \cite{Vogel2002,Bro1997}. Numerous experiments verified that enforcing a non-negativity constraint could lead to more accurate approximate solution\cite{Hanke2000,Nagy2000}.   We adopt {\bf Algorithm 1} to solve problems (\ref{Eq:min-increase-1}) and (\ref{Eq:min-decrease-1}).
\begin{algorithm}[h!]
\caption{GMM for EIT reconstruction}
\begin{itemize}
 \item[i.] Initialization: Set $\boldsymbol\sigma_L^{t_0}=\boldsymbol\sigma_0$ at time $t=t_0$ and solve forward problem (\ref{NeumannBVP}) to get reference solution $u_0$ and reference boundary data $\mathbb{V}_L(t_0)$. Compute sensitivity matrix $\mathbb{S}$ with reference solution $u_0$.
 \item[ii.] For time $t=t_n$, $n=1,2,\ldots$
 \begin{itemize}
 \item[1.] Read measured boundary voltage data ${\bf V}(t_n)$
 \item[2.]  Apply band-pass filter to get ventilation-induced signal ${\Bbb V}_L(t_n)$ as well as the time difference data ${\Bbb V}_L(t_n)-{\Bbb V}_L(t_{n-1})$
 \item[3.] Evaluate the change of $\sum_{j,k}\widetilde V_L^{j,k}$  from (\ref{Eq:observation})
     \begin{itemize}
         \item[(a)] If $\sum_{j,k}\widetilde V_L^{j,k}(t_n)\leq\sum_{j,k}\widetilde V_L^{j,k}(t_{n-1})$,  solve the problem  (\ref{Eq:min-increase-1}).
         \item[(b)] If $\sum_{j,k}\widetilde V_L^{j,k}(t_n)>\sum_{j,k}\widetilde V_L^{j,k}(t_{n-1})$,  solve the problem  (\ref{Eq:min-decrease-1}).
     \end{itemize}
 \item[4.] Output: For certain given $n\geq1$, $\boldsymbol\sigma_L^{t_{n}}$ solve problems (\ref{Eq:min-increase-1}) and (\ref{Eq:min-decrease-1}).
 \end{itemize}
 \item[iii.] Stop when $n$  reaches the set time step.
 \end{itemize}
\end{algorithm}

To solve  (\ref{Eq:min-alpha}) and (\ref{Eq:min-beta}), we used similar algorithm as solving (\ref{Eq:min-increase-1}) and (\ref{Eq:min-decrease-1}) with minor modifications of steps ii. 3. (a) and ii. 3. (b) in {\bf Algorithm 1}. Precisely, the step ii. 3. (a) and ii. 3. (b) in  algorithm 1 should be adapted as that in {\bf Algorithm 2}.
\begin{algorithm}[h!]
\caption{LMM for EIT reconstruction}
\begin{itemize}
  \item[ii. 3. (a)] If $\sum_{j,k}\widetilde V_L^{j,k}(t_n)\leq\sum_{j,k}\widetilde V_L^{j,k}(t_{n-1})$,   compute $\boldsymbol\alpha$ by (\ref{Eq:evaluate-alpha}) and
      solve problem (\ref{Eq:min-alpha}).
         \item[ii. 3. (b)] If $\sum_{j,k}\widetilde V_L^{j,k}(t_n)>\sum_{j,k}\widetilde V_L^{j,k}(t_{n-1})$, compute $\boldsymbol\beta$ by (\ref{Eq:evaluate-beta}) and
             solve problem (\ref{Eq:min-beta}).
\end{itemize}
\end{algorithm}

By adding monotonicity constraints of the conductivity change, the inverse problem will get better posed and the solution of (\ref{Eq:min-increase-1}), (\ref{Eq:min-decrease-1}) and (\ref{Eq:min-alpha}), (\ref{Eq:min-beta}) will provide more accurate and stable images compare with the standard LM. Numerical, phantom and human experiments results will be shown in the following  sections.

\section{Numerical simulations}
In this section, we show various  numerical experiments to test the proposed GMM and LMM and compare the results with that of the standard LM.

\subsection{2D numerical example}
\begin{figure}[h!]
\centering
\begin{tabular}{cc}
\includegraphics[width=7cm]{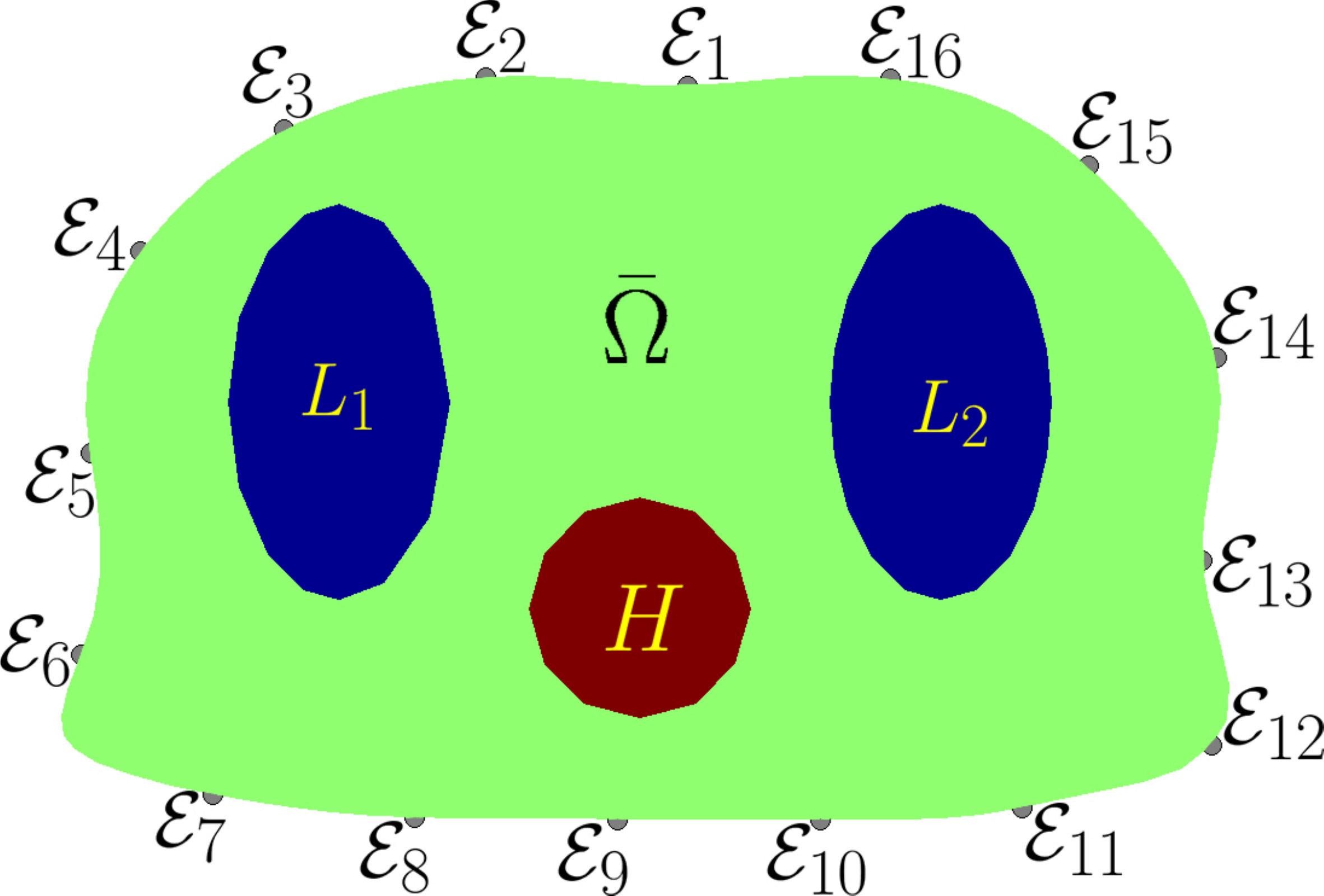}&
\includegraphics[width=7cm]{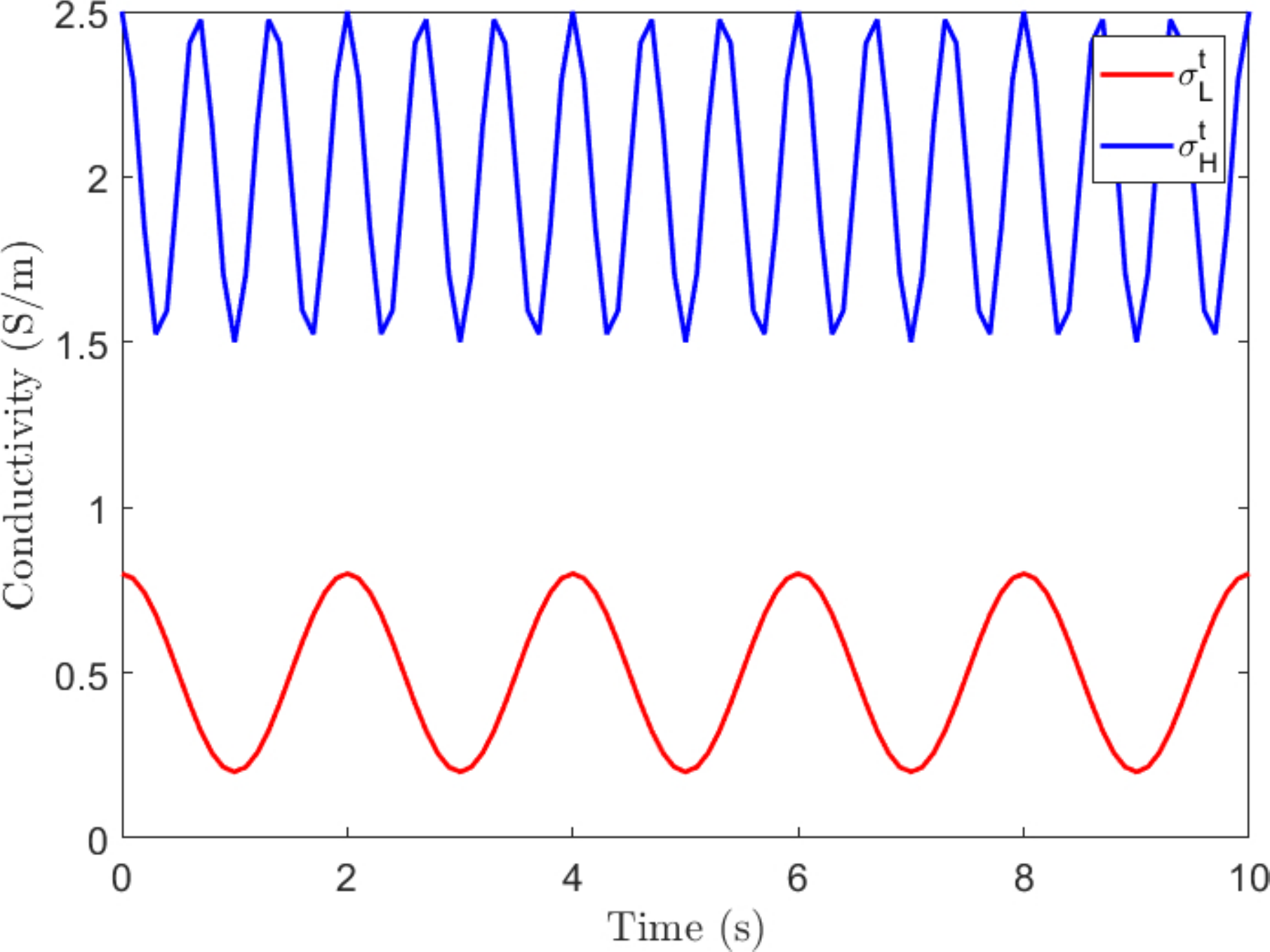}\\
(a) & (b)
\end{tabular}
\caption{{ Simulated 16-channel 2D EIT model. (a) the model geometry and electrodes configuration; (b) the simulated conductivity change in the lungs and heart.}}
 \label{Fig-2D-model}
\end{figure}

Numerical validation for the proposed monotonicity-based reconstruction methods were performed on a chest model occupying two dimensional domain $\bar\Om$. See Fig. \ref{Fig-2D-model}.   Three elliptic perturbations were added, one heart modeled in the region $H=\{(x,y)| (x-3.7)^2+(y-1.7)^2\leq 0.7^2\}$  and two lungs modeled in the regions $L_1=\{(x,y)| \f{(x-1.8)^2}{0.7^2}+\f{(y-3)^2}{1.25^2}\leq 1\}$ and $L_2=\{(x,y)| \f{(x-5.6)^2}{0.7^2}+\f{(y-3)^2}{1.25^2}\leq 1\}$.
The  conductivity  in the regions of $H, L_1$ and $L_2$  were set to be time dependent; $\sigma^t_H= 2+0.5\cos(3\pi t)$ in $H$  and  $\sigma^t_L= 0.5+0.3\cos(\pi t)$ in $L_1\cup L_2$. The background conductivity is $1$.

Finite element mesh with 766 triangular elements and 452 nodes was generated. Boundary voltages were simulated using 16-channel EIT system. There are  a total of 256 measurements each time frame by injecting currents and measuring boundary voltage adjacently. For simplicity, the contact impedances were ignored in the forward model.

\begin{figure}[h!]
\centering
\begin{tabular}{cc}
\includegraphics[width=7cm]{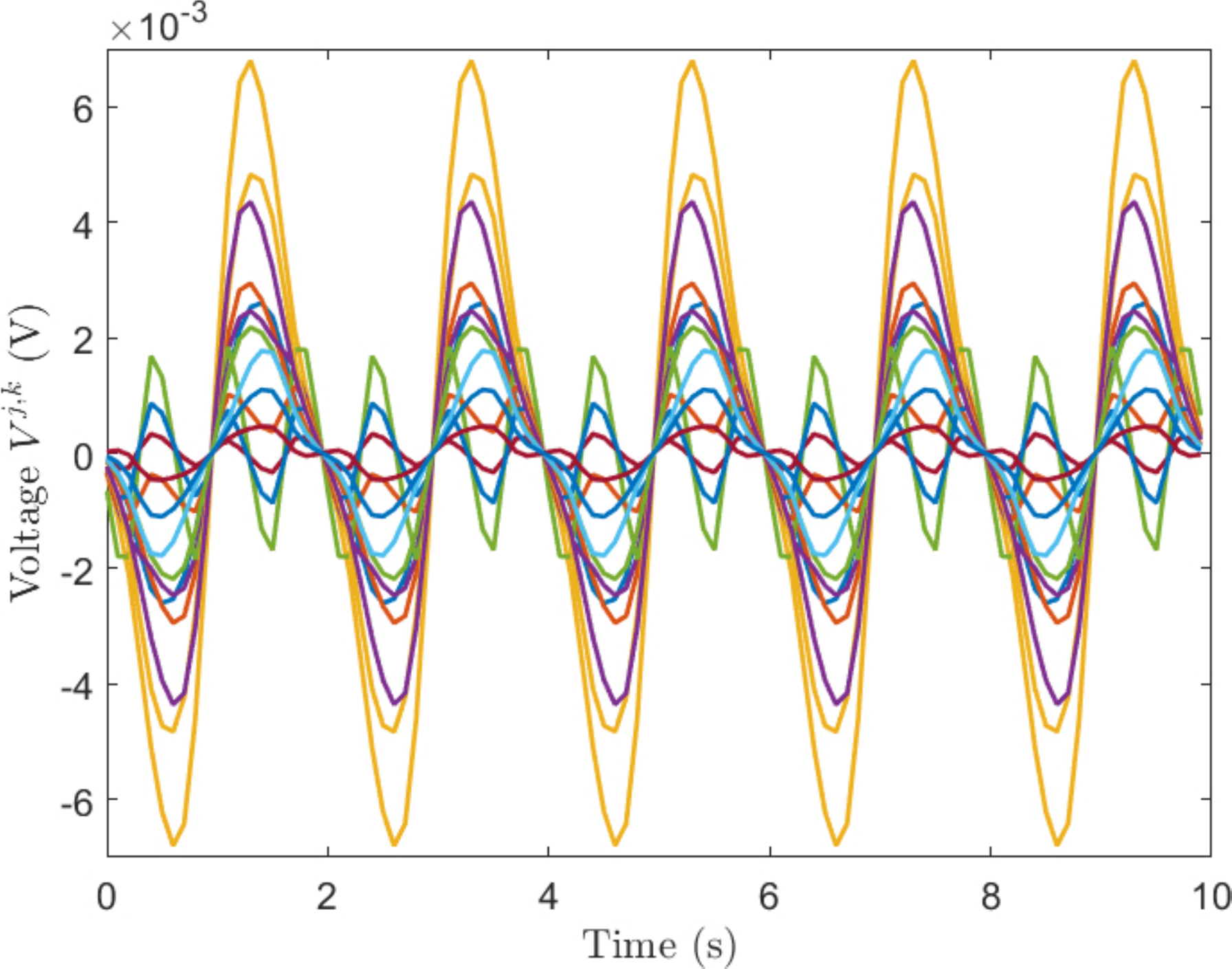}&
\includegraphics[width=7cm]{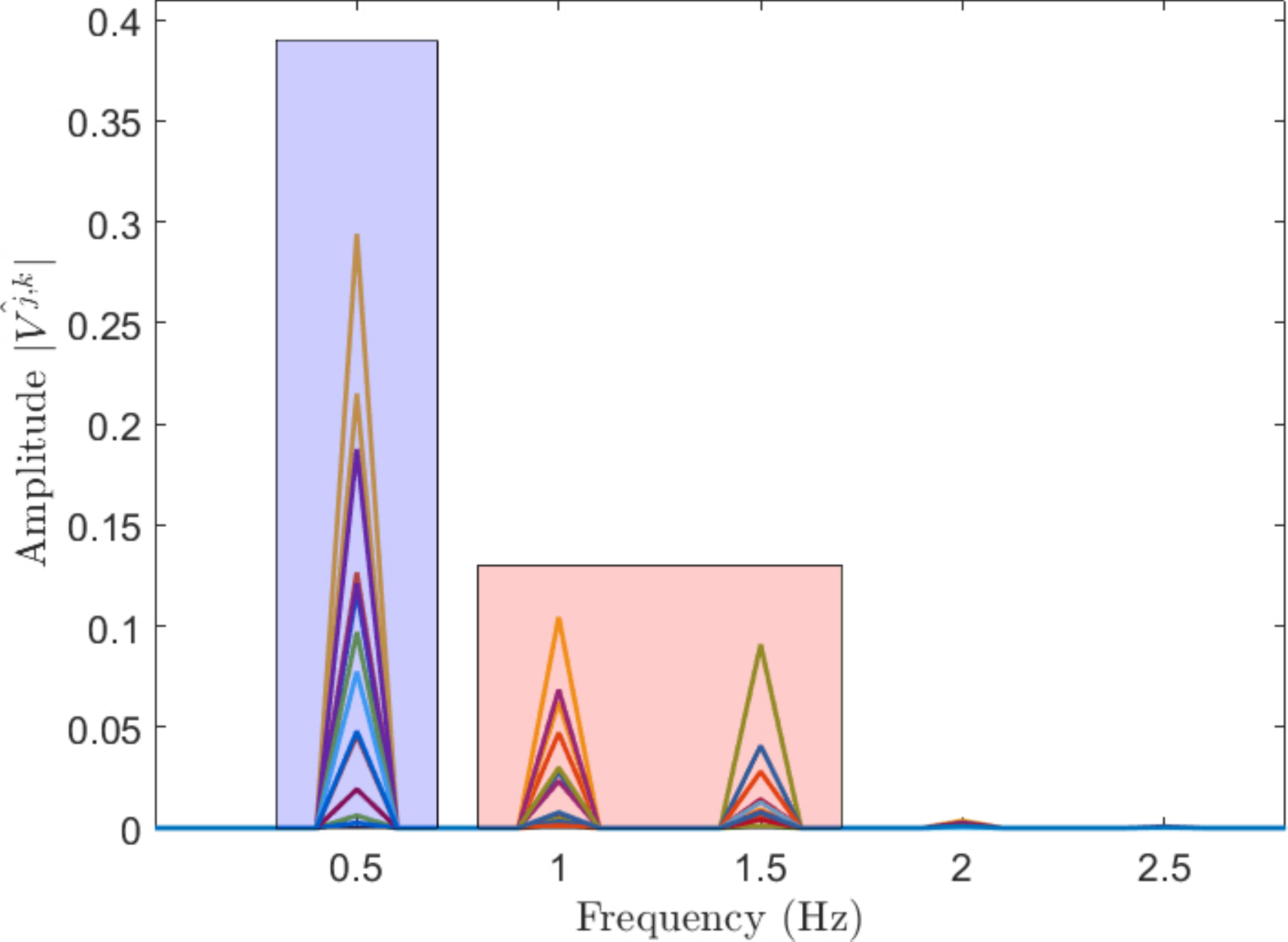}\\
(a) & (b)\\
\includegraphics[width=7cm]{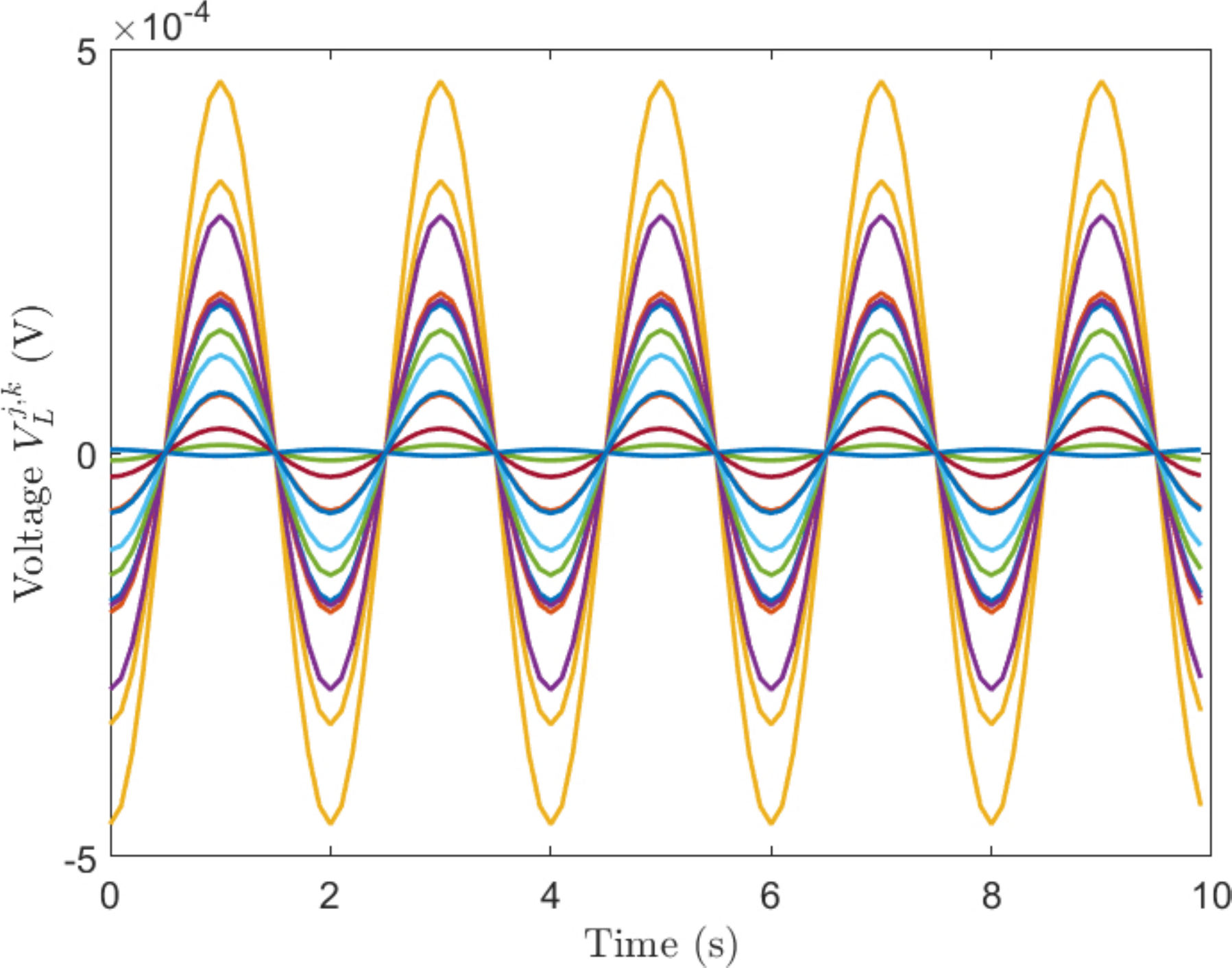}&
\includegraphics[width=7cm]{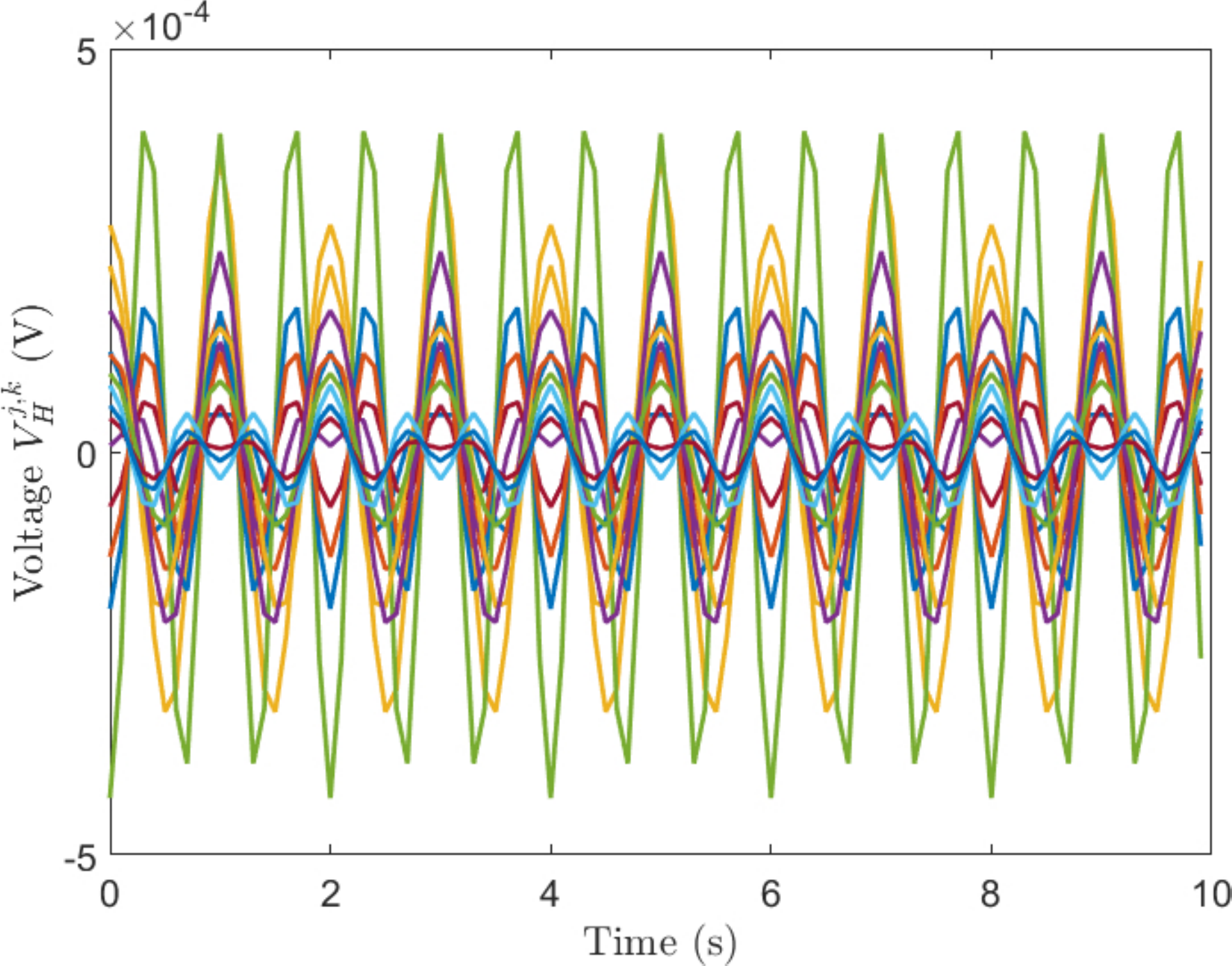}\\
(c)&(d)
\end{tabular}
 \caption{{ Simulated data separation. (a) the measured data $V^{j,k}$; (b) the Fourier transform $|\hat V^{j,k}|$; (c) the extracted ventilation induced data $V^{j,k}_L$ corresponding to frequency range in the blue shadow region of (b); (d) the high frequency data $V^{j,k}_H$ corresponding to frequency range in the pink shadow region of (b).}}
 \label{Fig-data-separation}
\end{figure}

{ With the conductivity shown in Fig. \ref{Fig-2D-model} (b), the simulated boundary data ${\bf V}$ is separated into two parts by applying band-pass filter, one part ${\bf V}_L$ is related to $\sigma^t_L$  and the other part ${\bf V}_H$ is related to  $\sigma^t_H$, see Fig. \ref{Fig-data-separation}. The cutoff frequency for ${\bf V}_L$ is 0.3 Hz  and 0.7 Hz, while the cutoff frequency for ${\bf V}_H$ is 0.8 Hz and 1.7 Hz. There are two peak frequency for ${\bf V}_H$ in Fig. \ref{Fig-data-separation} (b) which may be because of the nonlinearity of conductivity to voltage map. Fig. \ref{Fig-data-separation} shows that ventilation-induced data can be filtered out due to the nature of frequencies distinction between pulmonary and cardiac activities.}

\begin{figure}[h!]
\center
\includegraphics[width=14cm]{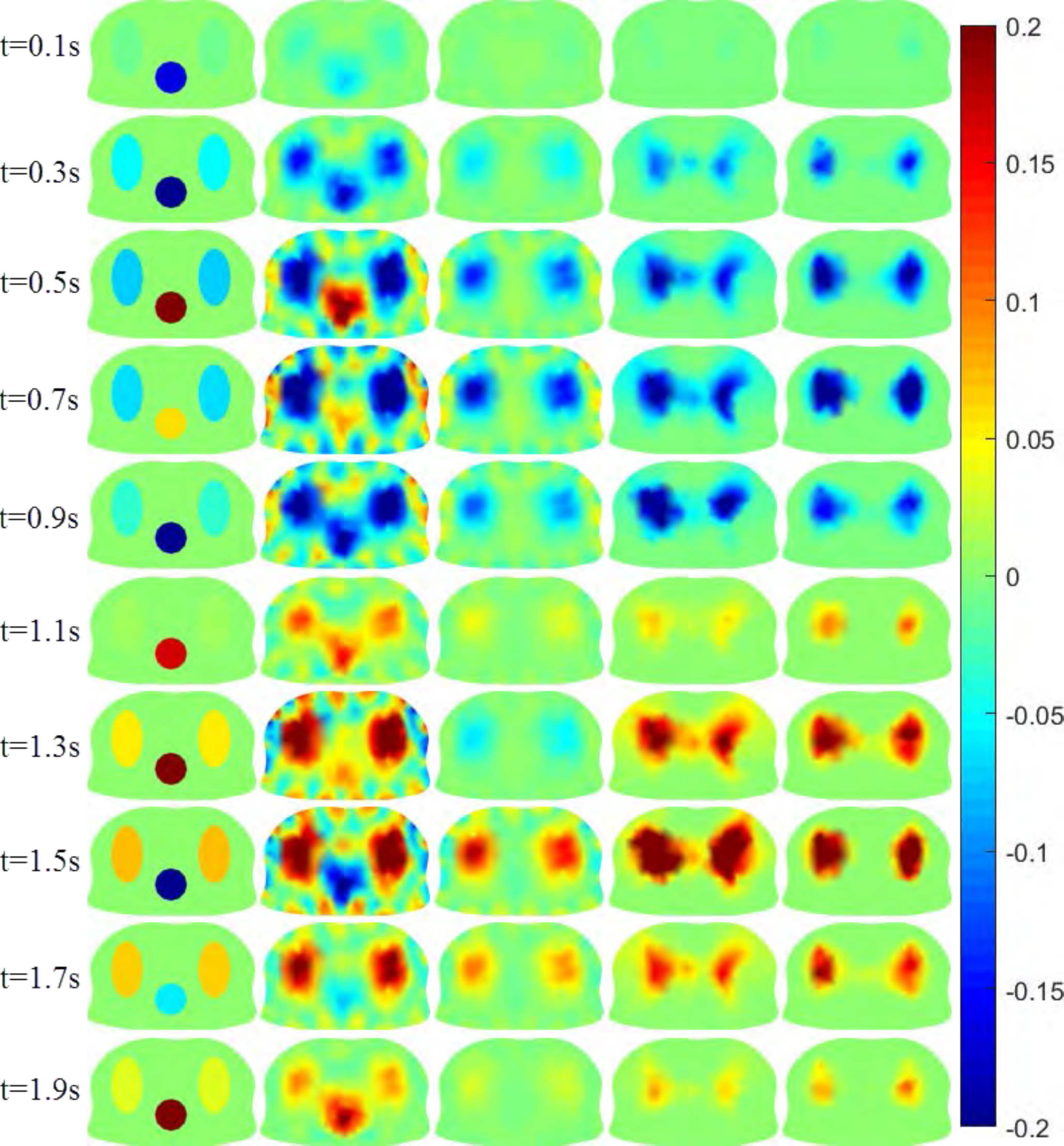}
\caption{{ Time-difference conductivity images in  the simulated  2D model. Rows indicate 10 time frames and columns indicate imaging methods. (1st col.) the true distribution of conductivity change; (2nd col.) the results using the standard LM with the unfiltered data ${\bf V}$; (3rd col.) the results using the standard LM  with the filtered data ${\bf V}_L$;
 (4th col.) the results using  GMM;  (5th col.) the results using  LMM.}
}
\label{Fig:2D-results}
\end{figure}

\begin{figure}[!ht]
\center
\includegraphics[width=14cm]{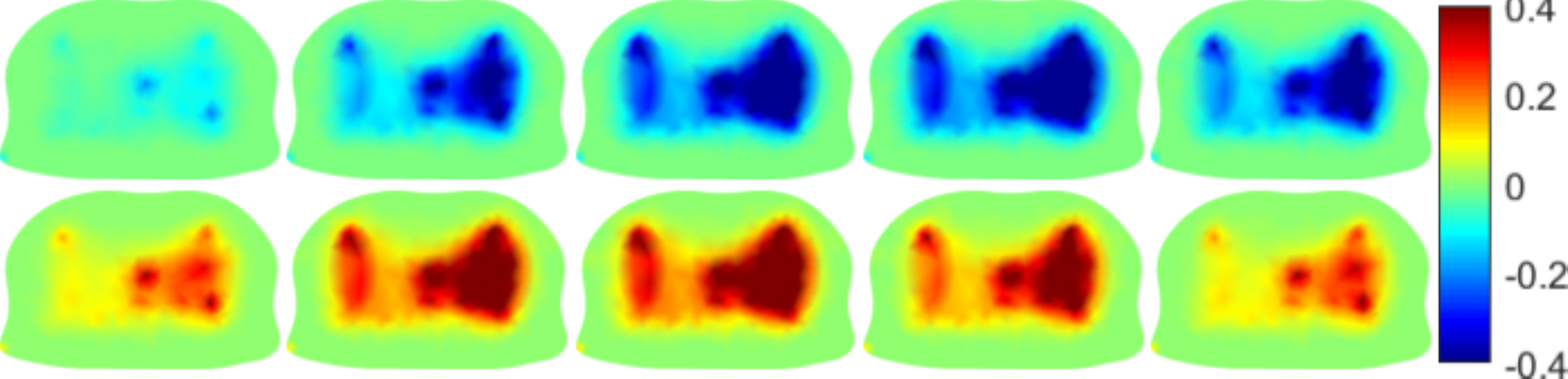}
\caption{{  The images in the 1st row are $-\boldsymbol\beta$ evaluated by  the rule (\ref{Eq:evaluate-beta}). The images in the 2nd row are $\boldsymbol\alpha$ evaluated by  the rule (\ref{Eq:evaluate-alpha}).}}
\label{Fig:2D-alpha}
\end{figure}

\begin{figure}[h!]
\center
\includegraphics[width=14cm]{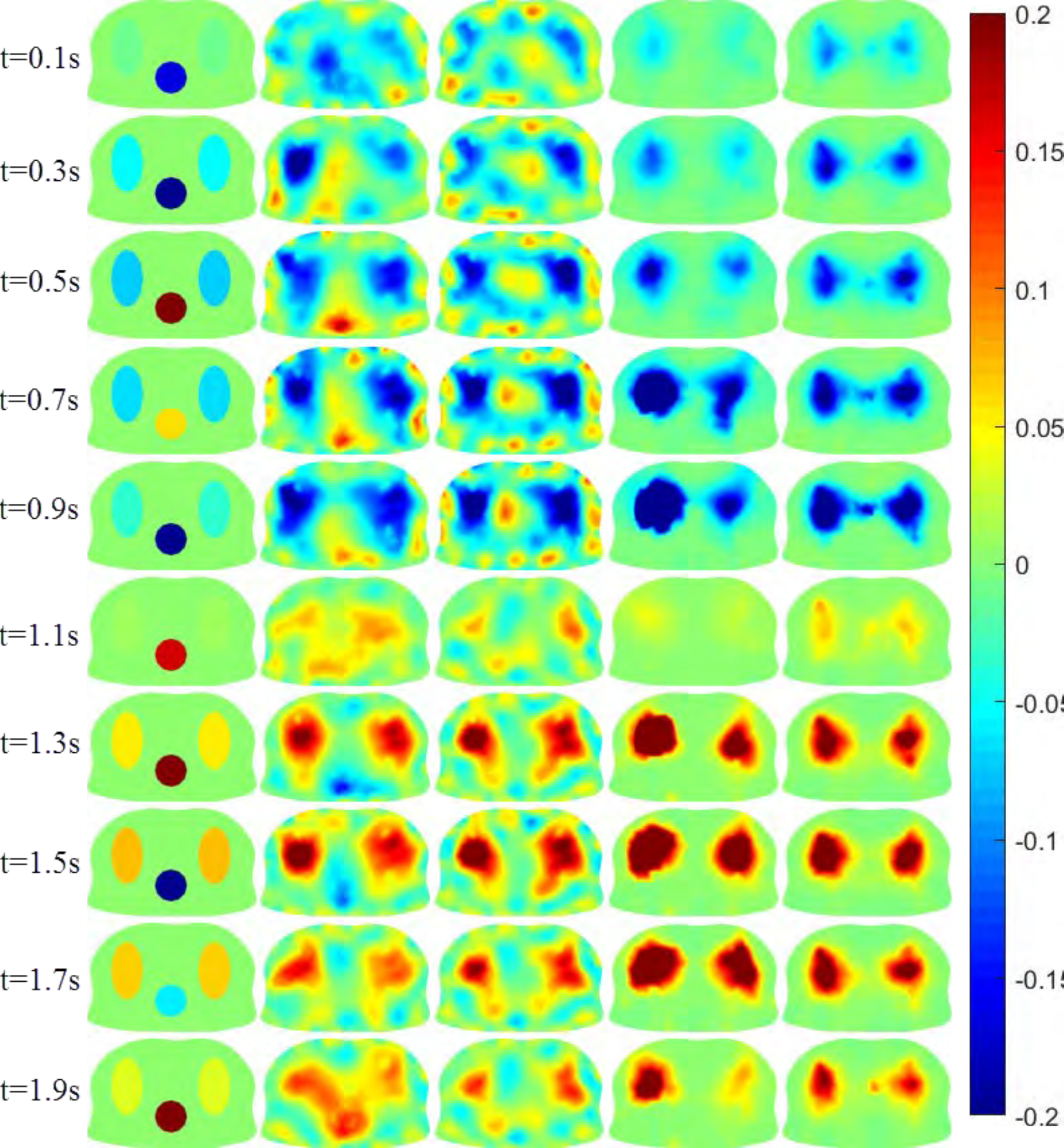}
\caption{{ Time-difference conductivity images in  the simulated  2D model with whiten noise (SNR=50dB).  Rows indicate 10 time frames and columns indicate imaging methods.  (1st col.) the true distribution of conductivity change;  (2nd col.) the results using the standard LM with the unfiltered data ${\bf V}$; (3rd col.) the results using the standard LM  with the filtered data ${\bf V}_L$;
 (4th col.)  the results using  GMM;  (5th col.) the results using  LMM.}
}
\label{Fig:2D-results_noise}
\end{figure}

  Fig. \ref{Fig:2D-results} shows reconstructed images at  ten time fames:  $t_1=0.1s, 0.3s, \ldots, 1.9s$. The first column of Fig. \ref{Fig:2D-results} shows the true conductivity distributions. The red color in the images indicates the increasing conductivity or positive values while blue color indicates  the decreasing conductivity or negative values. The second column shows EIT images using the standard LM with the unfiltered data ${\bf V}$. The third column shows EIT reconstructions using the standard LM with the filtered  data ${\bf V}_L$.
 The fourth row shows EIT images using GMM.   The fifth row displays images using LMM.  Fig. \ref{Fig:2D-alpha} shows images of $\boldsymbol\alpha$ and $-\boldsymbol\beta$ which are used for LMM in Fig. \ref{Fig:2D-results}. It shows clearly that $\boldsymbol\alpha$ and  $-\boldsymbol\beta$ provides the upper and lower bound of the conductivity change, respectively.  Moreover, numerical experiments with noisy data were conducted. Whiten noise was added to the voltage data to make the signal-to-noise ratio as 50dB. Fig. \ref{Fig:2D-results_noise} is the results with the noisy data.

  As shown in Fig. \ref{Fig:2D-results} and Fig. \ref{Fig:2D-results_noise},  results from the standard LM using unfiltered and filtered data have many artifacts. On the other hand, the proposed GMM and LMM provide reconstructions with less artifacts.  We chose best regularization $\lambda$ for the standard LM from many tests by varying $\lambda$. For the proposed GMM and LMM, very small $\lambda$ was chosen for these simulations and similarly for the subsequent simulations and experiments. Moreover, the performance of GMM and LMM are insensitive to the choice of $\lambda$ provided $\lambda$ is sufficiently small. However, the performance of the standard LM is highly depending on the choice of $\lambda$. We think the monotonicity constraints play an important role as a temporal regularization in the proposed methods.
\subsection{3D numerical example}
\begin{figure}[h!]
\centering
\includegraphics[width=8cm]{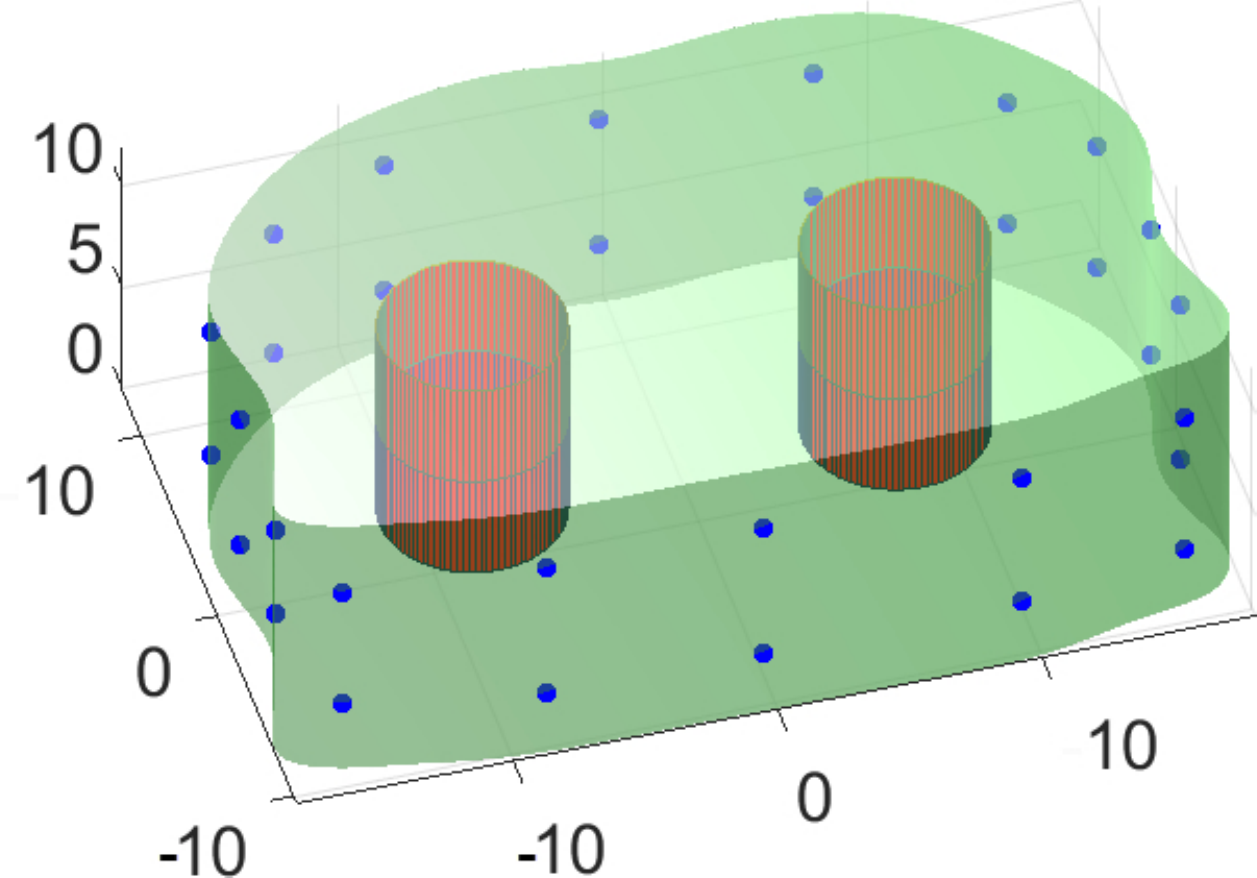}
\caption{Simulated 32-channel 3D EIT model and electrodes configuration.}
\label{Fig-3D-simulation}
\end{figure}

For 3D numerical simulation, thorax model was used as displayed in Fig. \ref{Fig-3D-simulation}.  To compute the forward solution,  4843 tetrahedra elements and 1195 nodes were used.   The thorax model $\bar\Om$ was contained in the cuboid $26cm\times17cm\times12cm$, and  32 electrodes were placed as shown in Fig. \ref{Fig-3D-simulation}.  For each pair of current injection through adjacent electrodes, there are 32 boundary voltage measurements between all pair of adjacent electrodes which lead to a total of 1024 measurements each time frame.  Inside the domain $\bar\Om$,  we  put two identical  cylinders whose size change with time. The size of cylinders at five different times $t_1, \cdots,  t_5$   are given in TABLE  \ref{Tb-numerical}. And the vertical center of all the cylinders are at $z=6cm$.    The conductivity value of these two cylinders  was set to be $0.5$ and background conductivity was $1$.
\begin{table}[h!]
\centering
\caption{Size of two identical  cylinders at five different times.}
\begin{tabular}{c||cccccc}
\hline
Time&$t_1$&$t_2$&$t_3$&$t_4$&$t_5$\\
\hline\\
Height& $6cm$&$7cm$&$9cm$&$7cm$&$6cm$\\
\hline\\
Radius& $1cm$&$2.5cm$&$3.5cm$&$2.5cm$&$1cm$\\
\hline
\end{tabular}
\label{Tb-numerical}
\end{table}

\begin{figure}[h!]
 \centering
\includegraphics[width=14cm]{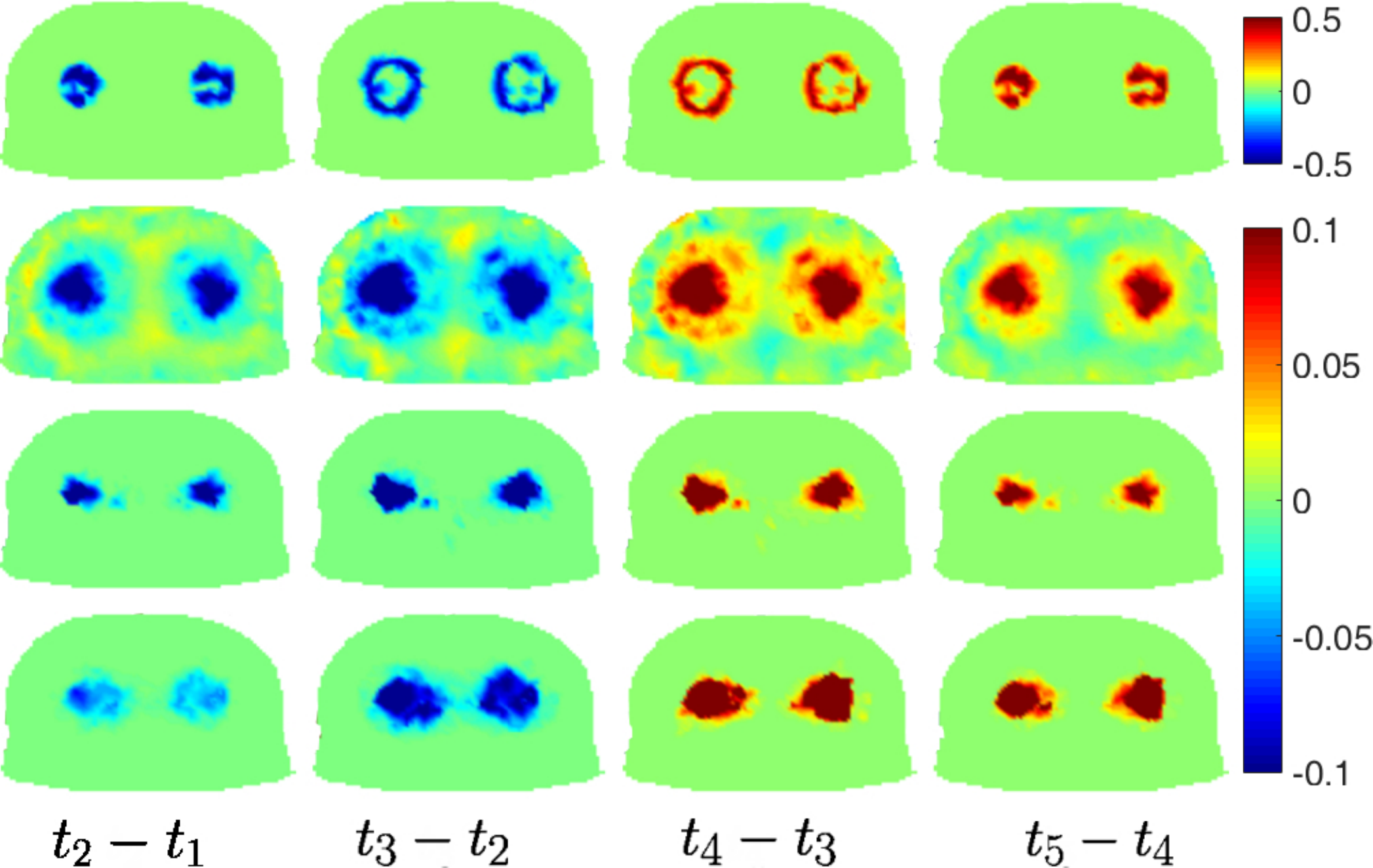}
\caption{Numerical experiment for time-difference conductivity imaging  at the slice $z=9cm$.  (1st row) true distribution of conductivity change; (2nd row) reconstructed images using the standard LM;  (3rd row)  reconstructed images using GMM; (4th row)  reconstructed images using  LMM.}
\label{Fig-3D-results-9}
\end{figure}

\begin{figure}[h!]
 \centering
\includegraphics[width=14cm]{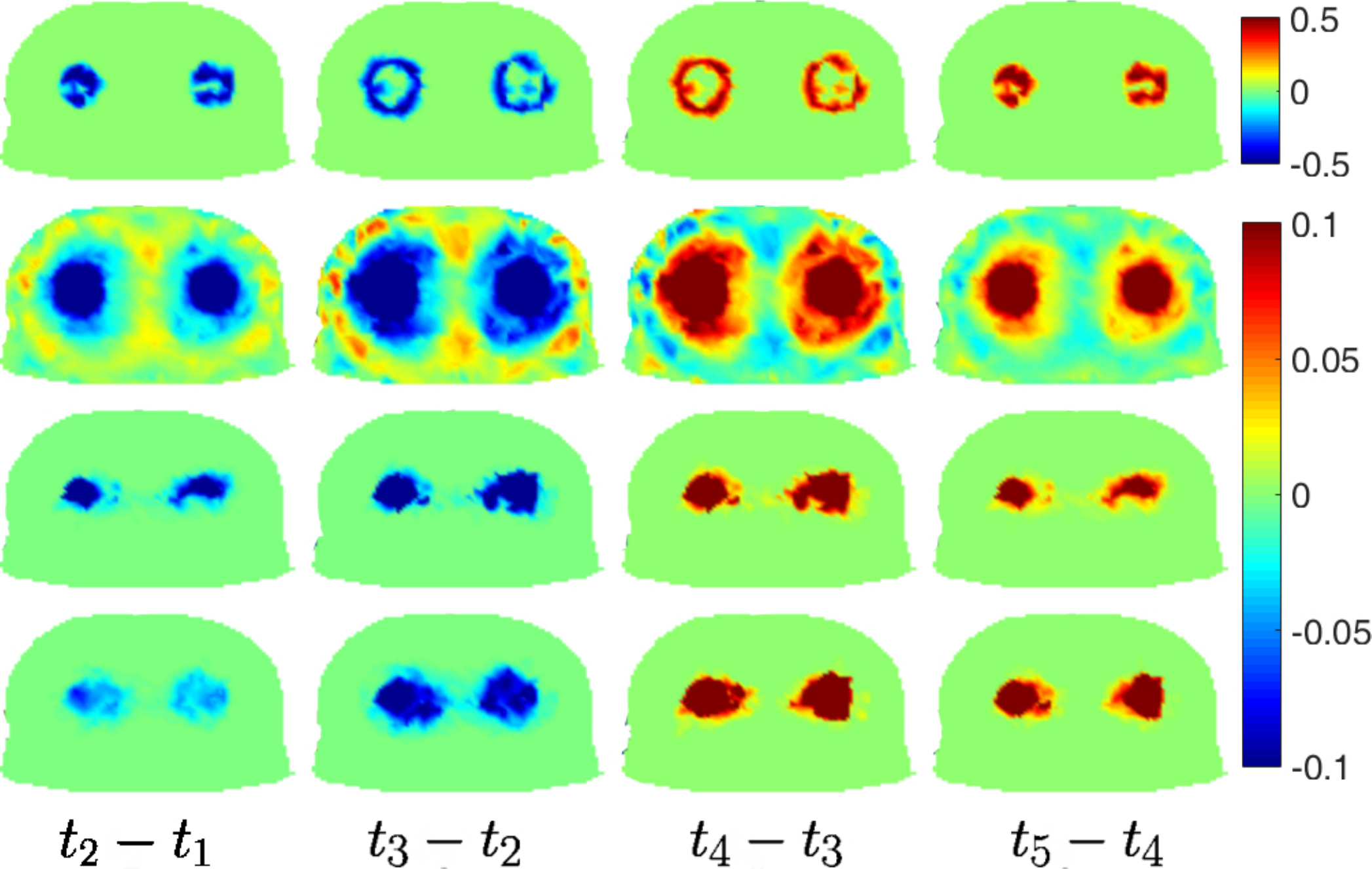}
\caption{Numerical experiment for time-difference conductivity imaging at the slice $z=3cm$.    (1st row) true distribution of conductivity change;  (2nd row) reconstructed images using the standard LM;  (3rd row)  reconstructed images using  GMM;  (4th row)  reconstructed images using  LMM.}
\label{Fig-3D-results-3}
\end{figure}

Fig. \ref{Fig-3D-results-9} and Fig. \ref{Fig-3D-results-3} show  EIT reconstructions of $xy$-slice images at $z=9cm$ and $z=3cm$,  respectively.
The first rows of Fig. \ref{Fig-3D-results-9}  and Fig. \ref{Fig-3D-results-3}  show the true distribution of  conductivity changes. The second rows  of Fig. \ref{Fig-3D-results-9}  and Fig. \ref{Fig-3D-results-3} show the reconstructed images using the standard LM.
The third rows  of Fig. \ref{Fig-3D-results-9}  and Fig. \ref{Fig-3D-results-3} show reconstructions using   GMM.
The fourth rows of Fig. \ref{Fig-3D-results-9}  and Fig. \ref{Fig-3D-results-3}  display the results using LMM.

\begin{figure}[ht!]
\centering
\includegraphics[width=14cm]{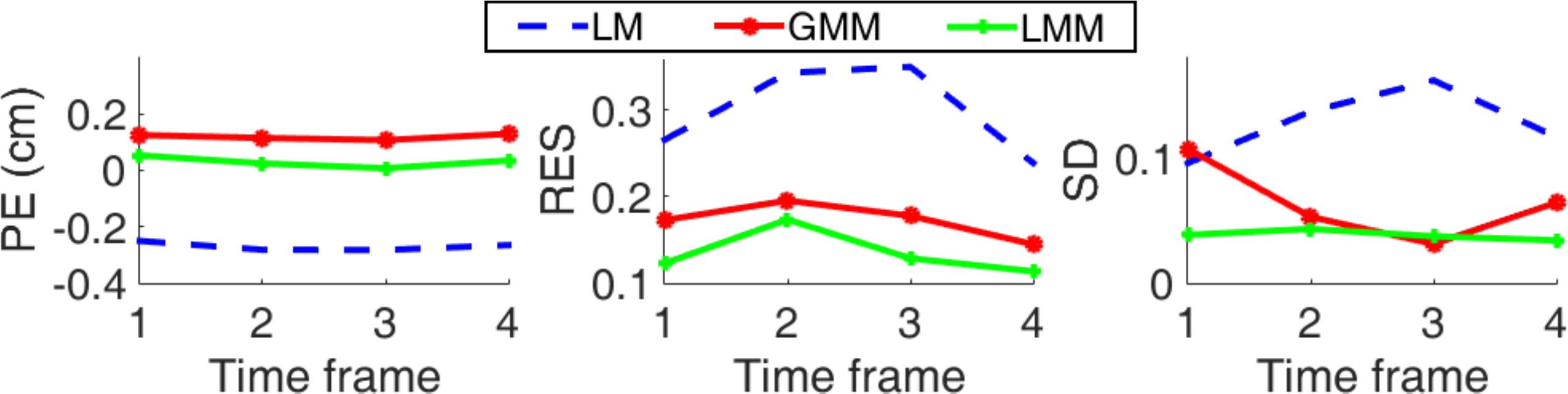}
\caption{Performance figures of merit \cite{Adler2009} for comparison of reconstructed images in Fig. \ref{Fig-3D-results-3}: position error (1st column), resolution (2nd column), shape deformation (3rd column).}
\label{Fig-quantitative83}
\end{figure}
 Fig. \ref{Fig-3D-results-9}  and  Fig. \ref{Fig-3D-results-3}  show that the reconstructions from the standard LM have many artifacts on the background while GMM and LMM have less artifacts. Moreover, the size of reconstructions from LMM are more accurate than that from LM and GMM.

For a quantitative analysis, we evaluated the performance figures of merit including the {  position error (PE, the smaller and less variable the better), resolution (RES, the smaller and more uniform the better ) and shape deformation (SD, the lower and more uniform the better)} of the reconstructions introduced in GREIT \cite{Adler2009}. Fig. \ref{Fig-quantitative83} shows PE, RES and SD of the results in Fig. \ref{Fig-3D-results-3}. The dashed blue lines,  red lines  and green lines show the results using the standard LM, GMM and LMM, respectively. These quantitative evaluations show that GMM and LMM have better performance than LM since the results from GMM and LMM have less position error, resolution and shape deformation. LMM performs slightly better than GMM as we see from Fig. \ref{Fig-3D-results-9}, Fig. \ref{Fig-3D-results-3} and Fig. \ref{Fig-quantitative83}.

\section{Phantom and human  experiments}
The proposed GMM and LMM were tested with phantom experiments. And for human experiments, results of GMM are presented. It is difficult to apply the  propose LMM to phantom and human experiments, because  the  tridiagonal elements of the data  matrix in (\ref{Eq:TimeData}) as well as in (\ref{Eq:LungData}) are not reliable due to  the unknown contact impedances of the current-driven electrodes. However, we include the reconstructions using LMM of phantom experiments to see how the results are affected by the unknown contact impedances of the current-driven electrodes.

\subsection{Phantom experiments}
\begin{figure}[h!]
\centering
\includegraphics[width=14cm]{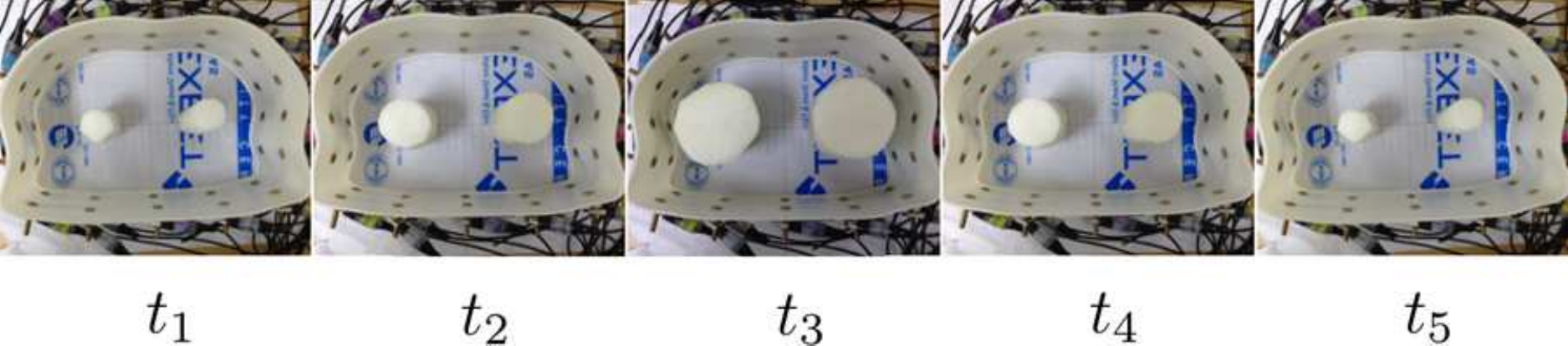}
\caption{32-channel EIT phantom.}
\label{Fig-Experiment}
\end{figure}

Experiments were conducted with thorax shape phantom using 32-channel Swisstom pioneer EIT system \cite{SPEIT}. To simulate the monotonically conductivity change of the lungs, five well trimmed radishes were used as shown in Fig.\ref{Fig-Experiment}.  The detailed information of the size of radishes are shown in TABLE  \ref{Tb-experiment}.  The size of thorax phantom is bounded by cuboid $26cm\times17cm\times12cm$ with two rings of electrodes. The phantom was filled with saline solution of conductivity $0.32 s/m$. 5mA current was injected  adjacently   at frequency 50kHz and the adjacent boundary voltage data was measured.

The Swisstom EIT system provides a full set of measurements,
$V^{j,k}(t)$, $j,k=1,\ldots,32$ for each time frame.  However,  the
voltage data on current-driven electrodes are known to be
prone to errors and affected by unknown contact impedances, so that
 only $32\times 29$ measurements ($V^{j,k}(t)$ with $|j-k|>1$)
are used for LM and GMM. { We applied the full data of $32\times 32$ measurements to LMM and find that the erroneous data from the current-driven electrodes can still provides acceptable results.}

\begin{table}[h!]
\centering
\caption{Size of radish for experiments at different times.}
\begin{tabular}{c||cccccc}
\hline
Time&$t_1$&$t_2$&$t_3$&$t_4$&$t_5$\\
\hline\\
Height& $9.5cm$&$10cm$&$11cm$&$10cm$&$9.5cm$\\
\hline\\
Radius& $1.5cm$&$2.2cm$&$3.5cm$&$2.2cm$&$1.5cm$\\
\hline
\end{tabular}

\label{Tb-experiment}
\end{table}

\begin{figure}[h!]
\centering
\includegraphics[width=14cm]{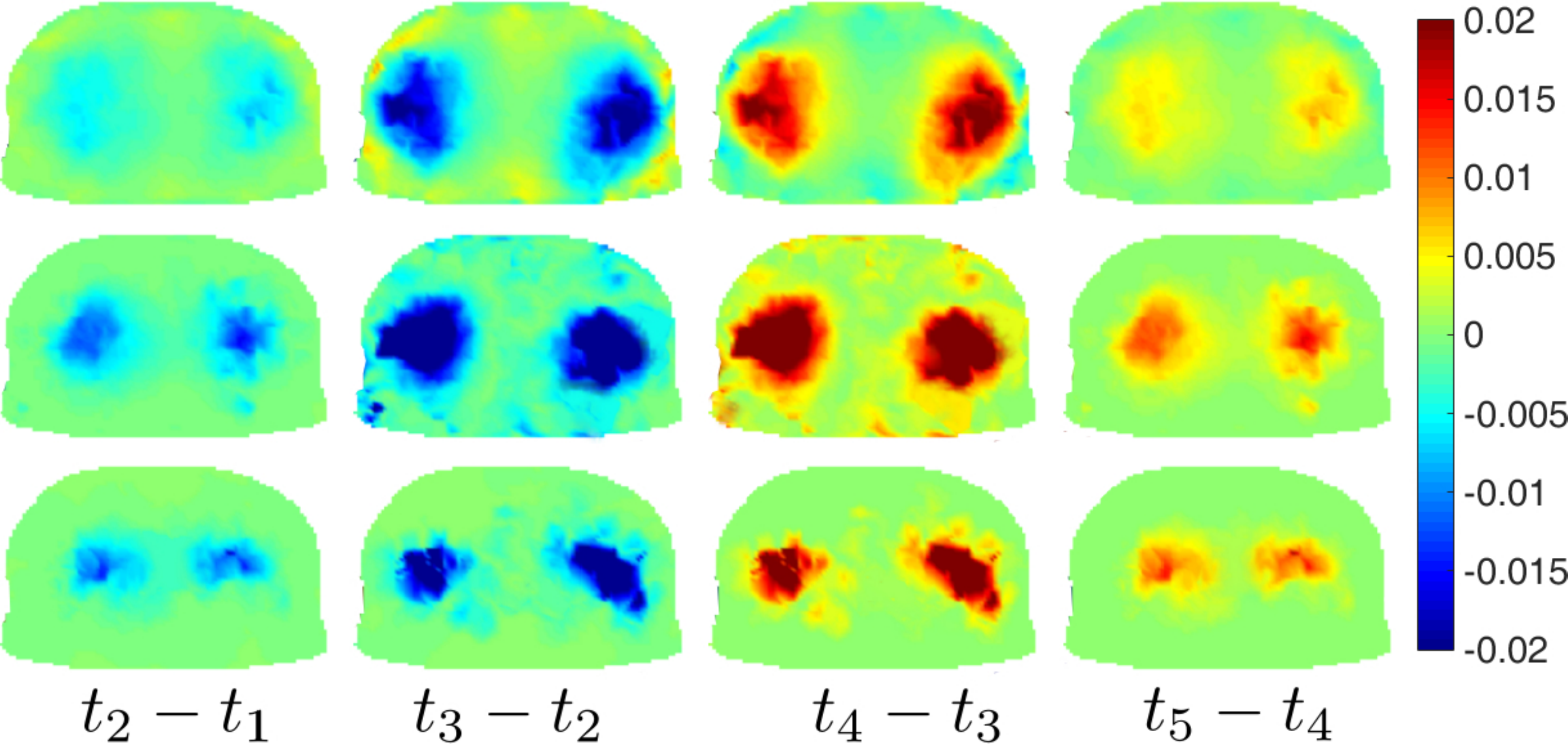}
\caption{Phantom experiment for time-difference conductivity imaging  at the slice $z=9cm$.  (1st row)  reconstructed images using the standard LM;  (2nd row)  reconstructed images using GMM; { (3rd row)  reconstructed images using  LMM.}}
\label{Fig-Experiment-9}
\end{figure}

\begin{figure}[h!]
\centering
\includegraphics[width=14cm]{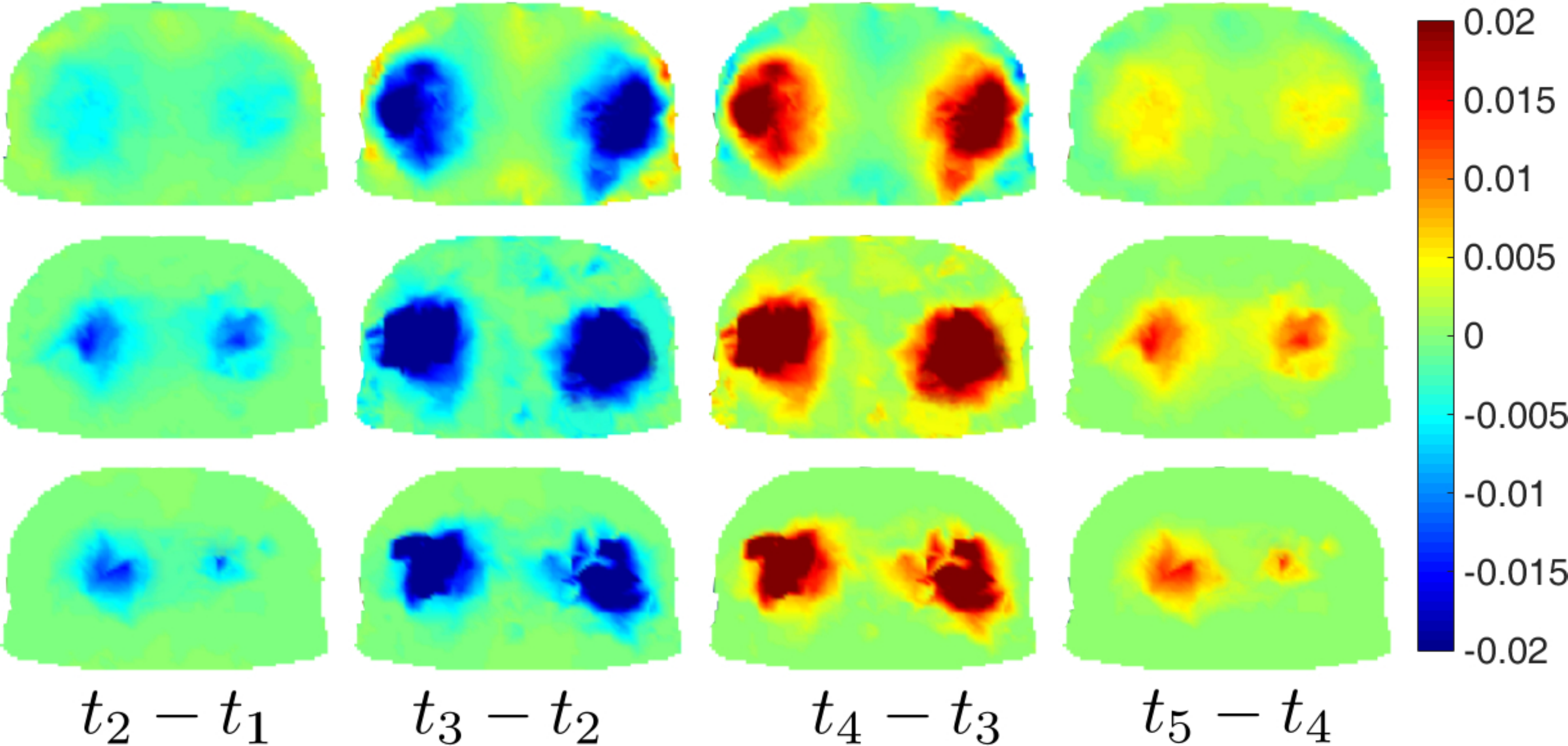}
\caption{Phantom experiment for time-difference conductivity imaging  at the slice $z=3cm$.     (1st row) reconstructed images using the standard LM;   (2nd row)  reconstructed images using GMM; { (3rd row)  reconstructed images using LMM.}}
\label{Fig-Experiment-3}
\end{figure}

Fig.\ref{Fig-Experiment-9} and Fig.\ref{Fig-Experiment-3}  display the  time difference conductivity images for phantom experiments  at slices $z=9cm$ and $z=3cm$, respectively.
The first rows of  Fig.\ref{Fig-Experiment-9} and Fig.\ref{Fig-Experiment-3}  show the reconstructed images using  the standard LM.
The second  rows of Fig.\ref{Fig-Experiment-9} and Fig.\ref{Fig-Experiment-3} show the reconstructed images using GMM. { The third  rows of Fig.\ref{Fig-Experiment-9} and Fig.\ref{Fig-Experiment-3} show the reconstructed images using LMM with full data.}

\begin{figure}[ht!]
\centering
\includegraphics[width=14cm]{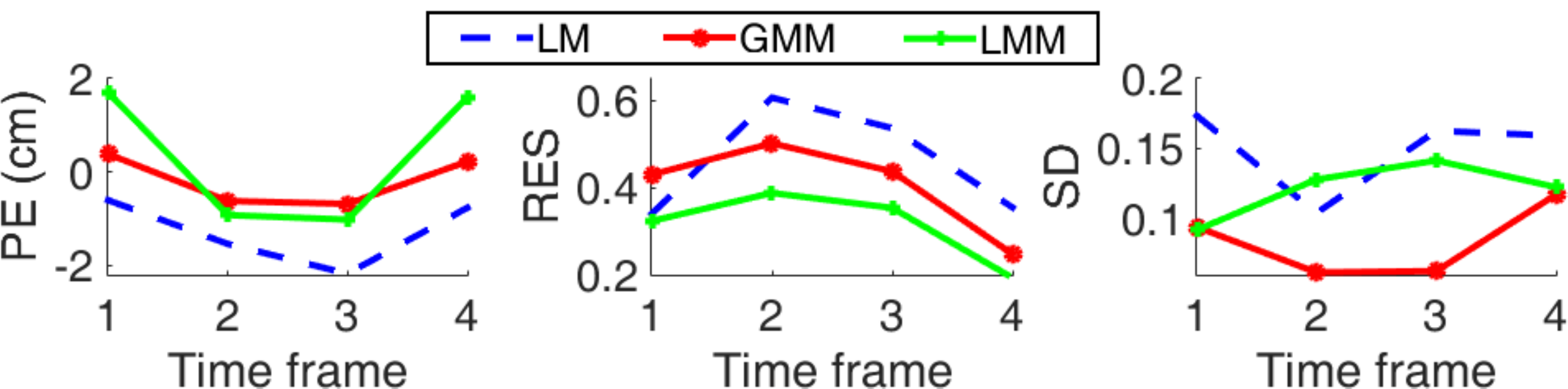}
\caption{ Performance figures of merit \cite{Adler2009} for comparison of reconstructed images in Fig. \ref{Fig-Experiment-3}: position error (1st column), resolution (2nd column), shape deformation (3rd column).}
\label{Fig-quantitative93}
\end{figure}
Fig.\ref{Fig-Experiment-9} and Fig.\ref{Fig-Experiment-3} show that even the results from  GMM  are not perfect,  they are relatively good compare with the results from the standard LM considering the shape and position of the objects.

Similar as the quantitative analysis for 3D numerical simulations in Fig. \ref{Fig-quantitative83}, the quantitative evaluation for the phantom experiments are shown in Fig. \ref{Fig-quantitative93}. We see that results from GMM have less position error and shape deformation. The results from LMM and LM have comparative position error and shape deformation.

Please note that the data measured on the current-driven electrodes are removed for the reconstruction of the standard LM and proposed GMM. Full data should be used for the proposed LMM since the LMM requires measurements on the current-driven electrodes. This is a preliminary
result to indicate that LMM will also be applicable to real data, but it is not comparable to GMM or LM as the voltage data on driven electrodes are being used for LMM.

We would like to emphasize that it is still an open problem to get reliable full measurements without being affected by the unknown contact impedances of current-driven electrodes. Solving this contact impedance problem is the topic of ongoing research which may rely on two approaches. One approach is the data interpolation using the measured data on the non-current-driven electrodes which is shown in a very recent work \cite{Harrach2015}. The other approach is to use the compound electrode \cite{Woo1992,Hua1993} with which different parts of electrode are used for the current injection and voltage measuring.

\subsection{Human experiments}
\begin{figure}[h!]
\centering
\begin{tabular}{cc}
\includegraphics[width=6cm]{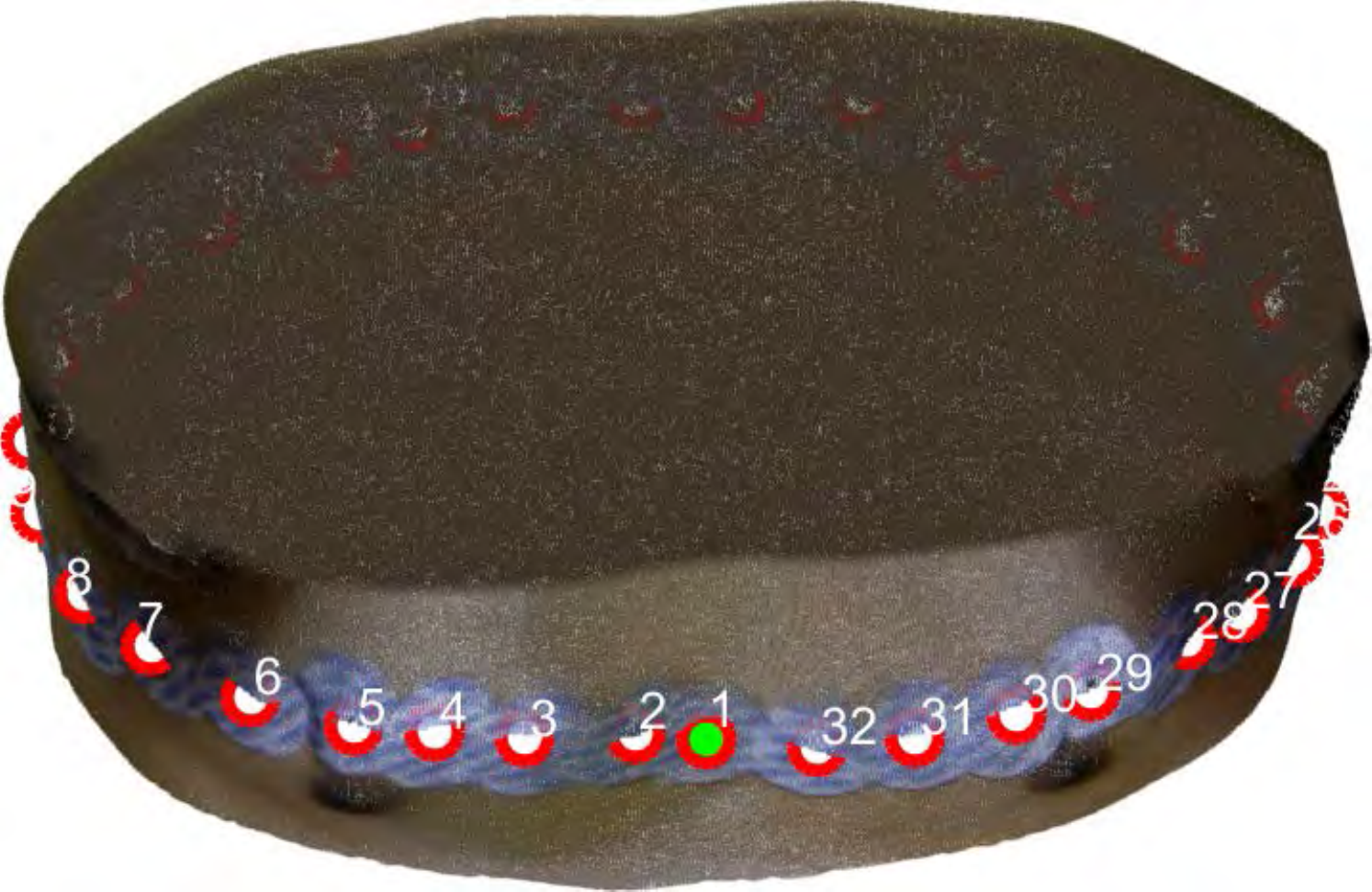}&
\includegraphics[width=6cm]{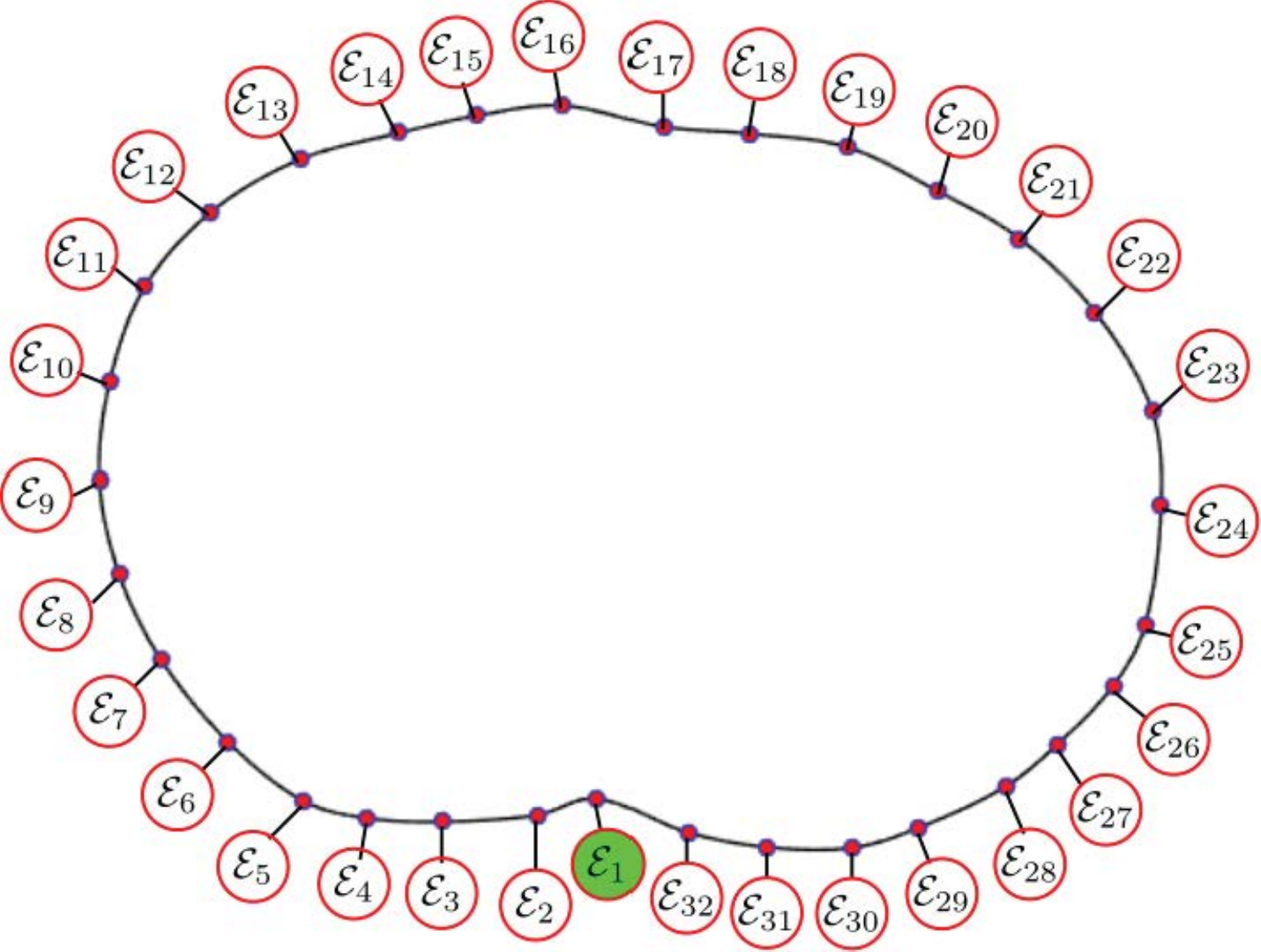}\\
(a) & (b)
\end{tabular}
\caption{ Automatic object  recognition to extract  boundary geometry and electrode positions for human experiment. (a) boundary geometry obtained by 3D scanner; (b) electrode positions.}
\label{Fig-3D}
\end{figure}

32-channel Swisstom BB$^2$ EIT system \cite{SBEIT} was used to conduct human experiments. A ring of 32 electrodes were attached to the thorax as shown in Fig. \ref{Fig-model-data} (a).  We used the Sense 3D scanner (Cubify 3D System) to scan the 3D geometry of the body. This 3D image with our Matlab GUI software allows us to extract the boundary geometry and  electrode positions. See Fig. \ref{Fig-3D} (a) and (b).  Current of 1mA at 150kHz was injected with skipping 4 electrodes, and boundary voltages were measured  at 10 frames per second. We applied GMM using a band-pass filtered data surrounding respiratory rate \cite{Grant2011}.  The source consistency observation in section \ref{Sub-source} was used  to determine the monotonic increase period and decrease period of conductivity.

\begin{figure}[h!]
\centering
\includegraphics[width=14cm]{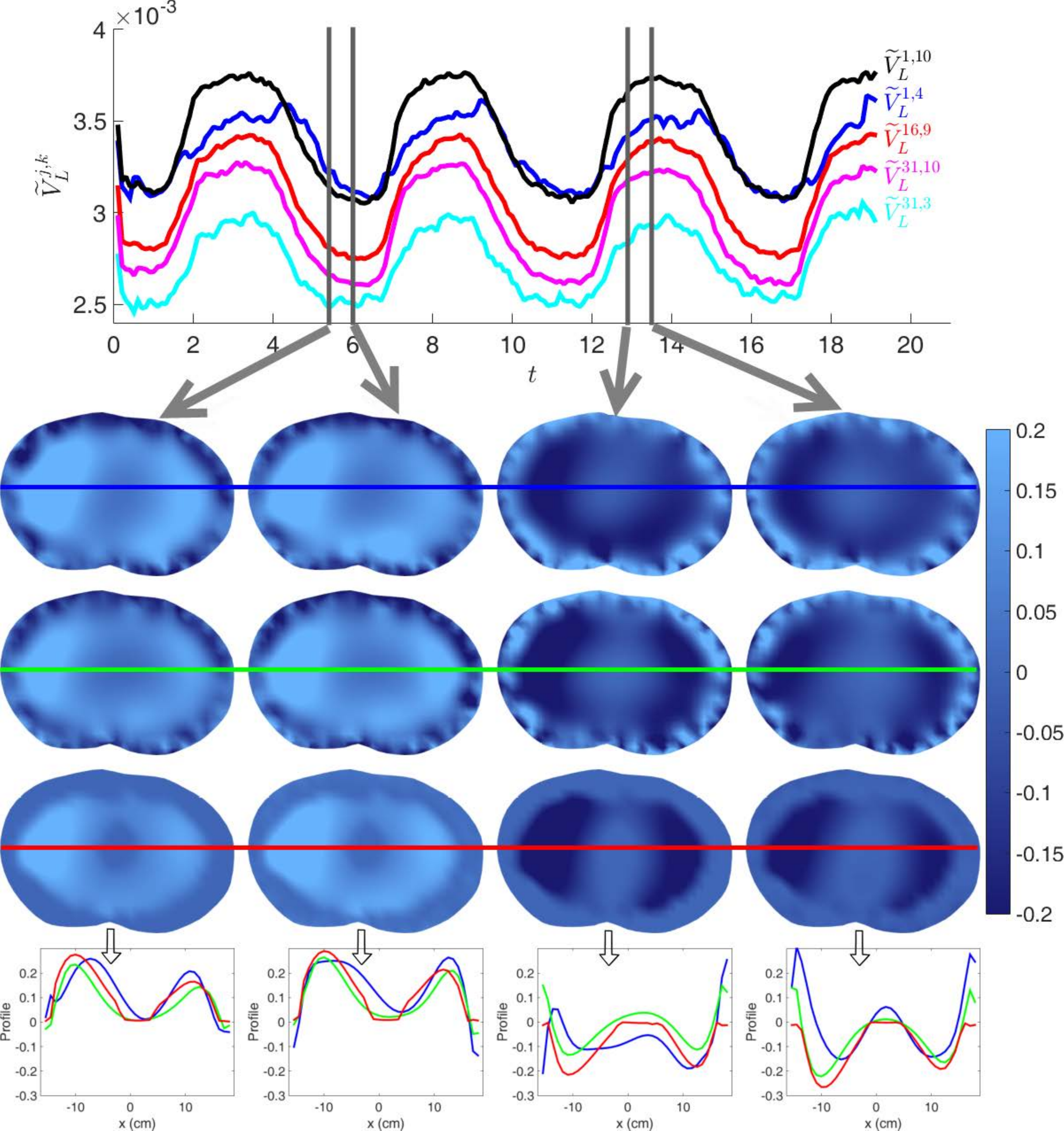}
\caption{Human experiment.  (1st row)  transformed voltages using (\ref{Eq:observation});  (2nd row)  time difference conductivity images using the standard LM with ${\bf V}$; (3rd row)   LM with ${\bf V}_L$; (4th row)  GMM;  (5th row)   their profiles.}
\label{Fig-Human-result}
\end{figure}

Fig. \ref{Fig-Human-result} shows the results using the standard LM and the  proposed GMM.  The first row of Fig. \ref{Fig-Human-result}  shows the transformed voltage. The second row  shows  the reconstruction using the standard LM with unfiltered data ${\bf V}$ and the third row shows the results from the standard LM with filtered data ${\bf V}_L$. The fourth row of Fig. \ref{Fig-Human-result} shows the results from GMM. The dark blue color indicates decease of conductivity while light blue color implies increase of conductivity in the second, third and fourth rows in Fig. \ref{Fig-Human-result}.  We compared the profiles of results from the standard LM and the proposed GMM in the last row of Fig. \ref{Fig-Human-result}.

Fig. \ref{Fig-Human-result} shows that results from GMM have less artifacts than that from the standard LM. The major advantage of  the proposed method is its robustness;   the reconstructed images are continuously depending on the measured data.
\section{Conclusions}
The proposed reconstruction methods allow lung EIT to visualize the monotonically varying conductivity distribution during inhalation and exhalation. It is based on the assumption that
lung conductivity monotonically decreases during inhalation (due
to the air flowing into the lungs) and monotonically increases during exhalation (due to the air leaving the lungs).
The periodicity of lung ventilation is used to extract its associated current-voltage data; all voltage differences
between electrodes increase during inhalation and
decrease during exhalation in terms of matrix definiteness regardless of the injection currents. This correlation between the time-differential of the current-voltage map and the changes of conductivity can be enforced in the reconstruction algorithm as a constraint.

We know that the inverse problem of EIT is ill-posed, and therefore any least square method by  data-fitting alone may not be able to provide useful images. This means that the boundary current-voltage data alone are insufficient to achieve robust reconstructions for making clinically useful
images.
Due to the inherent methodological limitation, the EIT reconstruction
algorithm requires a strategy that balances data fitting and a suitable regularization by imposing a certain constraint on the expected image. The proposed methods  use monotonicity as such a regularization.

Although EIT has limited resolution, its unique advantage lies in
its capability for continuous monitoring at the bedside.
Given the drawbacks of EIT (e.g., technical difficulties of the ill-posedness related to EIT data being insufficient to probe local conductivity changes), we need to
focus on a robust reconstruction method to allow this technique to provide indispensable information in clinical medicine.

\section*{Acknowledgment}
J.~K.~Seo and L.~Zhou were supported by the National Research Foundation of Korea (NRF) grant funded by the Korean government (MEST) (No. 2011-0028868, 2012R1A2A1A03670512).

\section*{Appendix A.}

\subsection*{A.1. Proof of the identities (\ref{Eq:T-data_standard}) and (\ref{Eq:T-data_standard_diff}).} \label{Apen-A}
{  Assume $\sigma^{t}(x)\in L^{\infty}_{+}(\Om)$ and $\f{\partial }{\partial t}\sigma^{t}(x)\in L^{\infty}(\Om)$, where $L^{\infty}_{+}(\Om)$ denotes the subspace of $L^{\infty}(\Om)$, functions with positive essential infima. By considering the shunt model which ignores the contact impedance between electrodes and the imaging domain, we prove the identities (\ref{Eq:T-data_standard}) and (\ref{Eq:T-data_standard_diff}) for all the $j,k=1,2,\cdots,E$.}

First note that
\begin{eqnarray*}
V^{j,k}(t)&=u_t^j|_{\mE_k}-u_t^j|_{\mE_{k+1}}=
\int_{\p\Om} u_t^j (\sigma^t\na u_{t}^k\cdot\n) ds= \int_\Om \sigma^t\nabla u_t^j \cdot \nabla u_t^k dx,
\end{eqnarray*}
which shows (\ref{Eq:T-data_standard}) and $V^{j,k}(t)=V^{k,j}(t)$.
With the same argument we obtain that
\begin{eqnarray*}
V^{j,k}(t)&=u_t^j|_{\mE_k}-u_t^j|_{\mE_{k+1}}=
\int_{\p\Om} u_t^j (\sigma^{t+\delta t}\na u_{t+\delta t}^k\cdot\n) ds= \int_\Om \sigma^{t+\delta t}\nabla u_t^j \cdot \nabla u_{t+\delta t}^k dx
\end{eqnarray*}
and
\begin{eqnarray*}
V^{j,k}(t+\delta t)&= V^{k,j}(t+\delta t)
= \int_\Om \sigma^{t} \nabla u_t^j \cdot \nabla u_{t+\delta t}^k dx.
\end{eqnarray*}
Hence,
\begin{eqnarray*}
\f{V^{j,k}(t+\delta t)-V^{j,k}(t)}{\delta t}=\int_\Om \f{\sigma^t-\sigma^{t+\delta t}}{\delta t} \nabla u_t^j \cdot{  \nabla u_{t+\delta t}^k }dx,
\end{eqnarray*}
{ and it follows from $\nabla u_{t+\delta t}^k\approx\f{\sigma^t}{\sigma^{t+\delta t}}\nabla u_{t}^k$ \cite{harrach2010factorization} and $\lim_{\delta t\rightarrow0}\f{\sigma^t}{\sigma^{t+\delta t}}=1$ that}
\begin{eqnarray*}
\f{d}{dt} V^{j,k}(t) = -\int_{\Om} \f{\p\sigma^t}{\p t} \na u_t^j\cdot\na u_t^k dx,
\end{eqnarray*}
which is the asserted identity (\ref{Eq:T-data_standard_diff}). \hfill $\Box$

\subsection*{A.2. Proof of estimate (\ref{Eq:estimate1}).}\label{Apen-B}

This type of monotonicity estimate goes back to \cite{Kang1997,Ikehata1998}, see also \cite{Harrach2010,Harrach2013,harrachresolution} for recent applications.

Let $a_1,\ldots,a_E\in \R$. For brevity, we write $u_t^a:=\sum_{j=1}^E a_j u_t^j$.
As in the proof of (\ref{Eq:T-data_standard}) and (\ref{Eq:T-data_standard_diff}) we have that
\begin{eqnarray*}
{\bf a}^T {\Bbb V}_L(t_{n-1}){\bf a}&= \int_\Om \sigma_L^{t_{n-1}}\nabla u_{t_{n-1}}^a \cdot \nabla u_{t_{n-1}}^a dx,\\
{\bf a}^T {\Bbb V}_L(t_n){\bf a}&= \int_\Om \sigma_L^{t_{n}}\nabla u_{t_{n}}^a \cdot \nabla u_{t_{n}}^a dx=\int_\Om \sigma_L^{t_{n-1}}\nabla u_{t_{n-1}}^a \cdot \nabla u_{t_{n}}^a dx.
\end{eqnarray*}
From
\begin{eqnarray*}
0& \leq \int_\Om \sigma^{t_{n-1}}_L \left| \nabla u_{t_{n-1}}^a - \nabla u_{t_n}^a \right|^2\\&= \int_\Om \sigma^{t_{n-1}}_L \left| \nabla u_{t_{n-1}}^a\right|^2
- 2  \int_\Om \sigma^{t_{n-1}}_L  \nabla u_{t_{n-1}}^a \cdot \nabla u_{t_n}^a
 \quad + \int_\Om \sigma^{t_{n-1}}_L \left| \nabla u_{t_n}^a \right|^2\\
& = {\bf a}^T \left( {\Bbb V}_L(t_{n-1}) - {\Bbb V}_L(t_n)\right) {\bf a}
+ \int_\Om (\sigma^{t_{n-1}}_L-\sigma^{t_{n}}_L) \left| \nabla u_{t_n}^a \right|^2,
\end{eqnarray*}
it follows that
\begin{eqnarray*}
{\bf a}^T \left(  {\Bbb V}_L(t_n) - {\Bbb V}_L(t_{n-1}) \right) {\bf a}
\leq \int_\Om (\sigma^{t_{n-1}}_L-\sigma^{t_{n}}_L) \left| \nabla u_{t_n}^a \right|^2,
\end{eqnarray*}
which is the first inequality in (\ref{Eq:estimate1}). The second inequality in (\ref{Eq:estimate1})
follows from interchanging $t_{n}$ and $t_{n-1}$.


\section*{References}
\begin{thebibliography}{99}
\bibitem{Adler1996Imag} Adler A and Guardo R 1996 Electrical impedance tomography: regularized imaging and contrast detection  {\it IEEE Trans. Med. Imag.} 15 170-179
\bibitem{Adler1996Eng} Adler A, Guardo R and Berthiaume Y 1996 Impedance imaging of lung ventilation: do we need to account for chest expansion? {\it IEEE Trans. Biomed. Eng.} 43 414-420
\bibitem{Adler2009} Adler A, Arnold J, Bayford R, Borsic A, Brown B, Dixon P, Faes T, Frerichs I, Gagnon H, Garber Y, Grychtol B, Hahn G, Lionheart W, Malik A,  Patterson R, Stocks J, Tizzard A, Weiler N and Wolf G 2009 GREIT: A unified approach to 2D linear EIT reconstruction of lung images {\it Physiol. Meas.} 30 S35 -S55
\bibitem{Ammari2009} Ammari H 2009 Mathematical modeling in biomedical imaging I: electrical and ultrasound tomographies, anomaly detection, and brain imaging {\it Mathematical Biosciences Subseries} 1983 Springer-Verlag Berlin Heidelberg
\bibitem{Barber1984a} Barber D C, Brown B H and Freeston I 1984 Imaging Spatial distributions of resistivity using applied potential tomography-APT {\it Information Processing in Medical Imaging} 446-462
\bibitem{Barber1984b} Barber D C and Brown B H 1984 Applied potential tomography  {\it J. Phys. E: Sci. Instrum. } 17 723-733
\bibitem{BHHM17} Barth A,  Harrach B, Hyv$\ddot{\mbox{o}}$nen N and Mustonen L 2017 Detecting stochastic inclusions in electrical impedance
tomography {\it Inverse Problems} 33 115012
\bibitem{Bikker2009} Bikker I, Leonhardt S, Bakker J and Gommers D 2009 Lung volume calculated from electrical impedance tomography in ICU patients at different PEEP levels  {\it Intensive Care Med.} 35 1362-1367
\bibitem{Borsic2010} Borsic A, Graham B M,  Adler A  and Lionheart W R 2010 In vivo impedance imaging with total variation regularization {\it IEEE Trans. Med. Imag.} 29 44-54

\bibitem{Bro1997}  Bro R and  Jong S D 1997 A fast non-negativity-constrained least squares algorithm {\it Journal of chemometrics}  11 393-401 
\bibitem{Cheney1999} Cheney M, Isaacson D  and Newell J 1999 Electrical impedance tomography  {\it SIAM review}  41 85–101 
\bibitem{Chiras2010}  Chiras D D 2010  Human biology  {\it Jones \& Bartlett Learning, 7th edition}, Chapter 8
\bibitem{choi2014regularizing}  Choi M K,  Harrach B and Seo J K 2014 Regularizing a linearized EIT reconstruction method using a sensitivity based factorization method  {\it Inverse Problems in Science and Engineering}  22  1029-1044 
\bibitem{Costa2008} Costa E,  Chaves C,  Gomes S,  Beraldo M,  Volpe S,  Tucci M,  Schettino I,  Bohm S,  Carvalho C,  Tanaka H,  Lima R and Amato M 2008 Real-time detection of pneumothorax using electrical impedance tomography  {\it Crit. Care Med.} 36  1230-1238 
\bibitem{Costa2009} Costa E,  Limab G and  Amatoa M2009 Electrical impedance tomography  {\it Curr. Opin. Crit. Care} 15 18-24 
\bibitem{Frerichs2009}  Frerichs I,  Pulletz S,  Elke G,  Reifferscheid F,  Schadler D,  Scholz J and  Weiler N 2009 Assessment of changes in distribution of lung perfusion by electrical impedance tomography  {\it Respiration}  77  282-291 
\bibitem{Gar17}  Garde H 2017 Comparison of linear and non-linear
monotononicity-based shape reconstruction using exact matrix
characterizations {\it Inverse Probl.  Sci.
Eng.} 26 1-18

\bibitem{GS17a} Garde H and Staboulis S 2017 Convergence and
regularization for monotonicitybased shape reconstruction in electrical
impedance tomography  {\it Numer. Math.} 135 1221–1251 

\bibitem{GS17b} Garde H and  Staboulis S 2017 The regularized monotonicity
method: detecting irregular indefinite inclusions  {\it arXiv preprint}
arXiv:1705.07372 

\bibitem{Grant2011}  Grant C A,  Pham T,  Hough J,  Riedel T,  Stocker C and  Schibler A 2011 Measurement of ventilation and cardiac related
    impedance changes with electrical impedance tomography  {\it Crit. Care}  15 R37
\bibitem{Hanke2000}  Hanke M,  Nagy J G and  Vogel C 2000 Quasi-Newton approach to nonnegative image restorations {\it Linear Algebra Appl.} 316  223-236 
\bibitem{hanke2011sampling}  Hanke M and   Kirsch A 2011 Sampling methods. In handbook of mathematical methods in imaging  {\it Springer, New York}  501-550 
\bibitem{Harrach2010} Harrach B and  Seo J K 2010 Exact shape-reconstruction by one-step linearization in electrical impedance tomography  {\it SIAM J. Math. Anal.}  42 1505-1518 
\bibitem{harrach2010factorization}   Harrach B,  Seo J K and  Woo E J 2010 Factorization method and its physical justification in frequency-difference electrical impedance tomography  {\it  IEEE Trans. Med. Imag.} 29  1918-1926 
\bibitem{harrach2013recent}  Harrach B 2013 Recent progress on the factorization method for electrical impedance tomography   {\it Comput. Math. Methods Med. } 2013 425184
\bibitem{Harrach2013}  Harrach B and  Ullrich M 2013 Monotonicity-based shape reconstruction in electrical impedance tomography  {\it SIAM J. Math. Anal.} 45  3382-3403 
\bibitem{harrachresolution}  Harrach B and  Ullrich M 2015 Resolution guarantees in electrical impedance tomography  {\it IEEE Trans. Med. Imag.} 34 1513-1521
\bibitem{Harrach2015}  Harrach B 2015 Interpolation of missing electrode data in electrical impedance tomography  {\it Inverse Problems}  31 115008
\bibitem{HLU15}  Harrach B,  Lee E and  Ullrich M 2015 Combining
frequency-difference and ultrasound modulated electrical impedance
tomography {\it Inverse Problems}  31  095003 

\bibitem{HM16a}  Harrach B and   Minh M N 2016 Enhancing residual-based
techniques with shape reconstruction features in electrical impedance
tomography {\it Inverse Problems}  32 125002 

\bibitem{HM16b}  Harrach B and  Minh M N 2018 Monotonicity-based
regularization for phantom experiment data in electrical impedance
tomography {\it Trends Math.}, accepted for publication.
\bibitem{Holder2005book}  Holder D 2005 Electrical impedance tomography: method, history and applications  {\it Series in Medical Physics and Biomedical Engineering}, CRC Press 
\bibitem{Hua1993}  Hua P,  Woo E J,  Webster J G and  Tompkins W J 1993 Using compound electrodes in electrical impedance tomography  {\it IEEE Trans. Biomed. Eng.} 40 29-34 
\bibitem{Ikehata1998}  Ikehata M 1998 Size estimation of inclusion  {\it J. Inverse Ill-Posed Probl.}  6  127-140 
\bibitem{Kang1997}  Kang H,  Seo J K and  Sheen D 1997 The inverse conductivity problem with one measurement: stability and estiation of size  {\it SIAM J. Math. Anal.}  28 1389-1405 
\bibitem{kirsch2008factorization}  Kirsch A and  Grinberg N 2008 The factorization method for inverse problems {\it Oxford University Press} 36 
\bibitem{Kunst2000}  Kunst P,  Vazquez A,  B$\ddot{\mbox{o}}$hm S,  Faes T,  Lachmann B,  Postmus P and  Vries P 2000 Monitoring of recruitment and derecruitment by electrical impedance tomography in a model of acute lung injury  {\it Crit. Care Med.}  28 3891-3895 
\bibitem{Leathard1994}  Leathard A D,  Brown B H,  Campbell J,  Zhang F,   Morice A H and   Tayler D 1994 A comparison of ventilatory and cardiac related changes in EIT images of normal human lungs and of lungs with pulmonary emboli  {\it Physiol. Meas.}  15  A137-A146 
\bibitem{Meier2008}  Meier T,  Luepschen H,  Karsten J,  Leibecke T,  Grossherr M,  Gehring H and  Leonhardt S 2008 Assessment of regional lung recruitment and derecruitment during a PEEP trial based on electrical impedance tomography  {\it Intensive Care Med.}  34  543-550 
\bibitem{Nagy2000}  Nagy J G and  Strakos Z 2000 Enforcing nonnegativity in image reconstruction algorithms  {\it In International Symposium on Optical Science and Technology}   International Society for Optics and Photonics  182-190 
\bibitem{Nopp1997}  Nopp P,  Harris N D,  Zhao T X and  Brown B H 1997 Model for the dielectric properties of human lung tissue against frequency and air content {\it  Med. \& Biol. Eng. \& Comput.} 35 695-702 
\bibitem{Oh2008} Oh T I,  Koo H,  Lee K H,  Kim S M,
     Lee J,  Kim S W,  Seo J K and  Woo E J 2008 Validation of a multi-frequency electrical impedance tomography (mfEIT) system KHU Mark1: impedance
    spectroscopy and time-difference imaging {\it Physiol. Meas.} 29 295-307 
\bibitem{Putensen2007}  Putensen C,  Wrigge H and  Zinserling J 2007 Electrical impedance tomography guided ventilation therapy  {\it Curr. Opin. Crit. Care} 13  344-350  
\bibitem{Seo2012book}  Seo J K and  Woo E J 2012 Nonlinear inverse problems in imaging  {\it Wiley, 1st edition}  December 
 \bibitem{Somersalo1992}  Somersalo E,  Cheney M and  Isaacson D 1992 Existence and uniqueness for electrode models for electric current computed tomography {\it SIAM J. Appl. Math}  52  1023-1040 
 \bibitem{Tamburrino2006}  Tamburrino A 2006 Monotonicity based imaging methods for elliptic and parabolic inverse problems {\it J. Inverse Ill-Posed Probl.} 14 633-642 
\bibitem{Tamburrino2002}  Tamburrino A and  Rubinacci G 2002 A new non-iterative inversion method for electrical resistance tomography  {\it Inverse Problems} 18   1809-1829 

\bibitem{Teschner}  Teschner E  and  Imhoff M 2011 Electrical impedance tomography: the realization of reginal ventilation monitoring  {\it Dr$\acute{a}$ger. Technology for Life} 
\bibitem{Vauhkonen1998} Vauhkonen M,   Vadasz D,  Karjalainen P A, E Somersalo E and  Kaipio J P 1998 Tikhonov regularization and prior information in electrical impedance tomography  {\it Tran. Med. Imag.} 17 285-293 
\bibitem{Vauhkonen1999}  Vauhkonen P J,   Vauhkonen M,   Savolainen T and  Kaipio J P 1999 Three-dimensional electrical impedance tomography based on the complete electrode model  {\it IEEE Trans. Biomed. Eng.} 46 1150-1160 
\bibitem{Vogel2002}  Vogel C R 2002 Computational methods for inverse problems  {\it SIAM}  Philadelphia, USA 
\bibitem{Wolf2013}  Wolf G,  Laberge C G,  Rettig J,  Vargas S,  Smallwood C,  Prabhu P,  Vitali S,  Zurakowski D and  Arnold J 2013 Mechanical ventilation guided by electrical impedance tomography in experimental acute lung injury  {\it Crit. Care Med.}  41 1296-1304 
\bibitem{Woo1992}  Woo E J,  Hua P,  Webster J G,  Tompkins W J and  Areny R P 1992 Skin impedance measurements using simple and compound electrodes  {\it Med. \& Biol. Eng. \& Comput.} 30  97-102 
\bibitem{SPEIT} Brochure for Swisstom EIT Pioneer Set, \url{http://www.swisstom.com/wp-content/uploads/Swisstom_brochure-PioneerSet_GB_1ST500-102_Rev002_web.pdf}, Swisstom AG, 2015
        \bibitem{SBEIT}Watch the lungs breathe!
EIT real-time monitoring with Swisstom BB$^2$, \url{http://www.swisstom.com/wp-content/uploads/SwisstomBB2_Brochure_GB_2ST100-112_Rev003.pdf}, Swisstom AG, 2015
\end{thebibliography}
\end{document}